\documentclass[twocolumn]{aastex631}
\usepackage{graphicx}
\usepackage[flushleft]{threeparttable}
\usepackage{tabularx}
\usepackage{booktabs}
\usepackage{amsmath}

\received{13 Mar 2024}
\revised{31 Oct 2024}
\accepted{05 Nov 2024}
\submitjournal{ApJ}

\shorttitle{Two $z\sim3.1$ Protoclusters}
\shortauthors{Nicandro Rosenthal et al.}
\graphicspath{{../figures/}{figures/}}

\begin{document}

\title{Spectroscopic Confirmation of A Massive Protocluster with Two Substructures at $\mathbf{z \simeq 3.1}$}

\correspondingauthor{Michael J. Nicandro Rosenthal}
\email{rosenthal@astro.wisc.edu}

\author[0000-0003-3910-6446]{Michael J. Nicandro Rosenthal}
\affiliation{Department of Astronomy, University of Wisconsin--Madison, 475 N. Charter Street, Madison, WI 53706, USA}

\author[0000-0002-3306-1606]{Amy J. Barger}
\affiliation{Department of Astronomy, University of Wisconsin--Madison, 475 N. Charter Street, Madison, WI 53706, USA}
\affiliation{Department of Physics and Astronomy, University of Hawaii, 2505 Correa Road, Honolulu, HI 96822, USA}
\affiliation{Institute for Astronomy, University of Hawaii, 2680 Woodlawn Drive, Honolulu, HI 96822, USA}

\author[0000-0002-6319-1575]{Lennox L. Cowie}
\affiliation{Institute for Astronomy, University of Hawaii, 2680 Woodlawn Drive, Honolulu, HI 96822, USA}

\author[0000-0002-1706-7370]{Logan H. Jones}
\affiliation{Department of Astronomy, University of Wisconsin--Madison, 475 N. Charter Street, Madison, WI 53706, USA}
\affiliation{Space Telescope Science Institute, 3700 San Martin Drive, Baltimore, MD 21218, USA}

\author[0000-0003-4248-6128]{Stephen J. McKay}
\affiliation{Department of Physics, University of Wisconsin--Madison, 1150 University Ave., Madison, WI 53706, USA}

\author[0000-0003-1282-7454]{Anthony J. Taylor}
\affiliation{Department of Astronomy, University of Wisconsin--Madison, 475 N. Charter Street, Madison, WI 53706, USA}
\affiliation{Department of Astronomy, The University of Texas at Austin, Austin, TX, USA}

 \begin{abstract}

\noindent
We present the results of a Keck and NOEMA spectroscopic survey of 507 galaxies, where we confirm the presence of two massive overdensities at $z = 3.090-3.110$ and $z = 3.133-3.155$ in the neighborhood of the GOODS-N, each with over a dozen spectroscopically confirmed members. We find that both of these have galaxy overdensities of NIR-detected galaxies of $\delta_{\rm gal,obs} = 6-9$ within corrected volumes of $(6-7)\times 10^3~{\rm cMpc}^3$. We estimate the properties of the $z = 0$ descendants of these overdensities using a spherical collapse model and find that both should virialize by $z \simeq 0.5-0.8$, with total masses of $M_{\rm tot} \simeq (6-7) \times 10^{14}~{\rm M}_\sun$. The same spherical collapse calculations, as well as a clustering-of-clusters statistical analysis, suggest a $>$80\% likelihood that the two overdensities will collapse into a single cluster with $M_{\rm tot} = (1.0-1.5) \times 10^{15}\,M_\sun$ by $z \sim 0.1-0.4$. The $z = 3.14$ substructure contains a core of four bright dusty star-forming galaxies with $\Sigma {\rm SFR} = 2700 \pm 700~{\rm M}_\sun~{\rm yr}^{-1}$ in a volume of only 280 cMpc$^{3}$.

 \end{abstract}
%TC:endignore


%%%%%%%%%%%%%%%%%%%
\section{Introduction} \label{sec:intro}
%%%%%%%%%%%%%%%%%%%
Galaxy clusters are the largest gravitationally bound structures in the Universe and have a pronounced influence on the evolution of galaxies contained within them. For decades, we have known that galaxies within denser environments, like groups and clusters, tend to be redder, more elliptical, and have lower star formation rates (SFRs) than field galaxies at similar redshifts \citep[e.g.,][]{Balogh98,Muzzin09}. This transition toward a ``red and dead'' galaxy population in dense environments has been shown to be the combined result both of secular feedback and environmental quenching.

At $z \gtrsim 1$, however, this SFR-density relation is inverted, with galaxies in denser environments being much more actively star-forming than those in co-eval field environments \citep{Elbaz07, Cooper08, Brodwin13}. This is believed to be at least in part due to time evolution: at early times, the most massive haloes accumulated gas faster in their deeper gravitational potential wells. This accumulation of gas allowed for an earlier onset of star formation, leading to increased SFRs at high redshifts. However, these overdense environments also quenched earlier, due to the relatively early onset of stellar, active galactic nuclei (AGNs), and environmental feedback, resulting in the red and dead population in galaxy clusters today. Understanding this population of protoclusters, the progenitors of today's galaxy clusters, is therefore essential to understand the buildup of mass in the Universe overall.

While large numbers of galaxy clusters have been discovered in wide-field, homogenous surveys out to $z \sim 1.5$ \citep[e.g.,][]{Bleem15, Balogh21}, far fewer $z \gtrsim 2$ protoclusters have been confirmed, with largely heterogenous selection \citep[e.g.,][and references therein]{Casey16, Alberts22}. Unlike nearby clusters, which can be identified by high concentrations of red galaxies \citep[e.g.,][]{Muzzin09}, or thermal X-rays \citep[e.g.,][]{Cruddace02, Stanford06}, or Sunyaev-Zeldovich effect signatures \citep[e.g.,][]{Bleem15} of the hot intracluster medium, protoclusters are not yet virialized \citep[][]{Chiang13, Muldrew15, Overzier16} and have bluer, more highly star-forming galaxies than the field. Protoclusters are therefore identified by the detection of large numbers of star-forming galaxies on  $\sim$10 cMpc scales \citep[e.g.,][and references therein]{Overzier16}, with the gold standard being large numbers of spectroscopic detections in a window of $\Delta z \approx 0.02-0.03$ \citep[e.g.,][]{Casey15, Cucciati18, Polletta21}. Due to the expense of spectroscopic observations at the highest redshifts, many $z \gtrsim 3$ protoclusters have only a few---or even a single---spectroscopically confirmed galaxy \citep[e.g.,][]{Wang21}, making it impossible to assess accurately their overdensities or kinematics, and thereby their masses. To date, there have only been a handful of $z \gtrsim 3$ protoclusters with $\gtrsim$10 spectroscopically confirmed members \citep[e.g.,][]{Steidel98, Miller18, Long20}. Enlarging this sample is therefore critical to understanding the evolution of overdense environments in the first 2 Gyr of the Universe.

Deep and unbiased spectroscopic surveys, like those conducted on nearby clusters, are generally too expensive to complete over large volumes at the high redshifts of protoclusters. Instead of using blank-field surveys, many protoclusters are identified by surveying the areas around rare, massive galaxies, such as dusty star-forming galaxies \citep[DSFGs; e.g.,][]{Casey16} or AGNs \citep[e.g.,][]{Ivison00}. Some protoclusters contain multiple DSFGs with individual SFRs $\gtrsim 100~{\rm M}_\sun~{\rm yr}^{-1}$ within volumes of only a few hundred cMpc$^3$ \citep[e.g.,][]{Miller18,Long20,Wang21}. These protocluster ``cores'' are the most densely star-forming extended structures in the Universe. 

\begin{figure*}[tbh]
    \centering
    \includegraphics[width=0.7\linewidth]{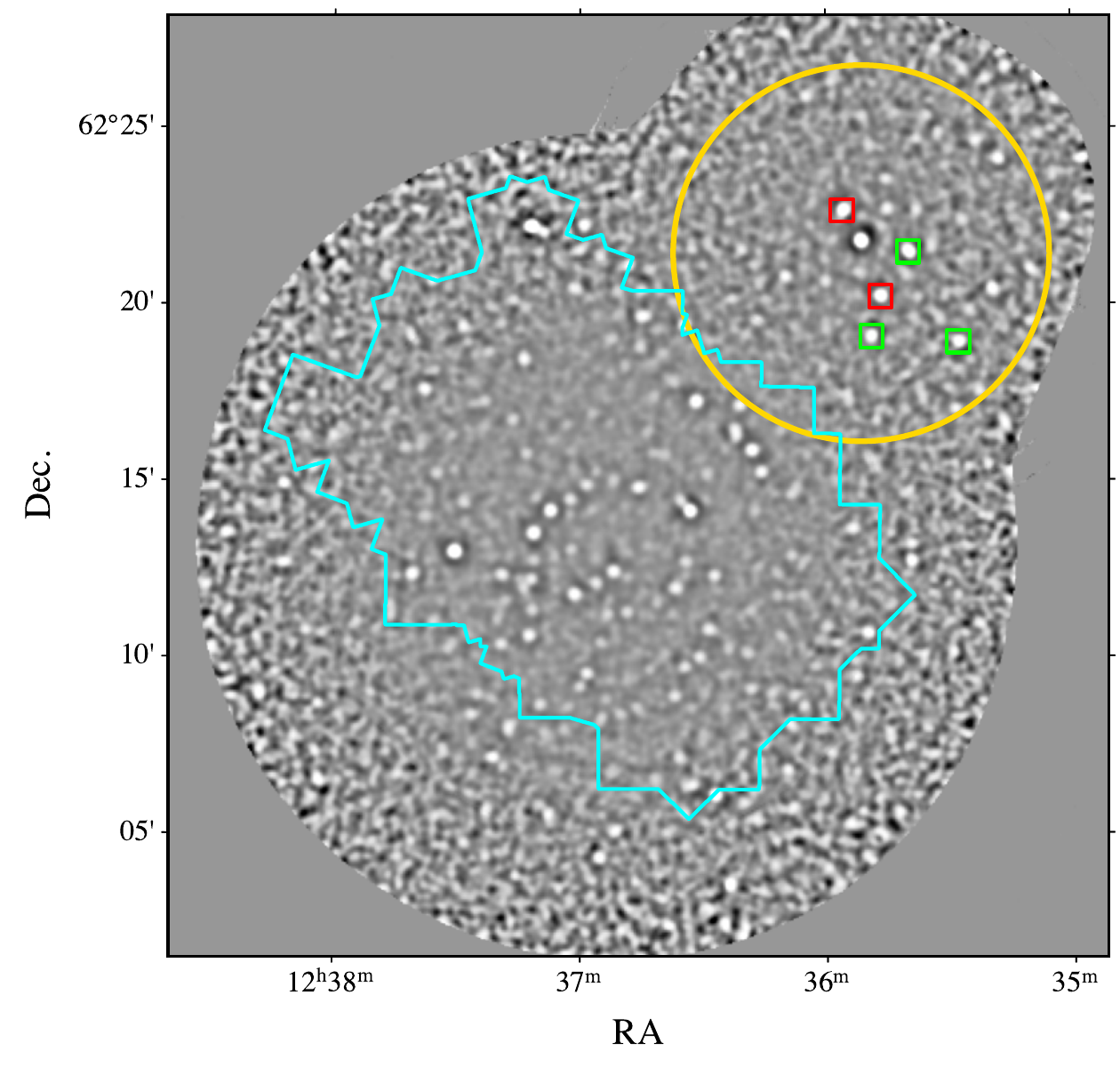}
    \caption{SCUBA-2 850\,$\mu$m map of the GOODS-N from SUPER GOODS \citep{Cowie17, Jones21}. The HST/WFC3 F160W footprint of CANDELS \citep[][]{Grogin11}, which roughly corresponds to the regions covered by the \citet{Barger08}, DEEP3 \citep[][]{Cooper11}, 3D-HST \citep[][]{Brammer12}, and MOSDEF \citep[][]{Kriek15} spectroscopic surveys, is marked in cyan. The two $z \sim 3.14$ DSFGs from \citet{Jones21} are marked by the red squares, and the gold circle is a 10 cMpc ($5.4'$) radius circle around the average position of those DSFGs, which is where we conducted our Keck observations. The three NOEMA targets described in Section \ref{subsec:noema} are marked by the green squares.}
    \label{fig:map}
\end{figure*}

Using 2\,mm and 3\,mm linescans with the Northern Extended Millimeter Array (NOEMA), \citet{Jones21} obtained CO spectroscopic redshifts (speczs) of three ultra-bright ($S_{\rm 850\mu m} \geq 11$ mJy) DSFGs at the very edge of the SCUBA-2 coverage of the GOODS-N. Two of the three were at $z \sim 3.14$ (red squares in Figure \ref{fig:map}) and separated by only $2.7'$, or 5.0 cMpc at $z = 3.14$. \citet{Jones21} surmised that these DSFGs may trace a massively star-forming protocluster at this redshift. However, due to their rarity, DSFGs are very biased tracers of mass \citep{Miller15}. Thus, confirming and characterizing a protocluster requires additional spectroscopic observations of other candidate protocluster members.

In this work, we present our spectroscopic surveys of galaxies near the $z \sim 3.14$ DSFGs from \citet{Jones21}. In Section \ref{sec:obs}, we describe our NOEMA observations of a further three SCUBA-2 sources (Section \ref{subsec:noema}), as well as our Keck/LRIS and Keck/MOSFIRE observations of ``normal'' star-forming galaxies (Section \ref{subsec:keck}). In Section \ref{sec:results}, we present the results of these observations, including the detection of two $z \sim 3.1$ structures. In Section \ref{sec:protoclusters}, we calculate the overdensities of these two structures and show that they will likely reach $z = 0$ masses of $M_{\rm tot} \gtrsim 5 \times 10^{14}~{\rm M}_\sun$. In Section \ref{sec:hyper}, we discuss the likely future of these overdensities and show that they are likely to combine to form a single cluster by $z = 0$, and in Section \ref{sec:core}, we note the existence of a dust-rich protocluster core in the $z = 3.14$ substructure. Finally, we summarize our work in Section \ref{sec:summary}.

We assume a concordance flat $\Lambda$CDM cosmology with $H_0 = 70~{\rm km~s^{-1}~Mpc^{-1}}$, $\Omega_{\rm m,0} = 0.3$, and $\Omega_\Lambda = 0.7$ throughout this work. All magnitudes are in the AB magnitude system.

%%%%%%%%%%%%%%%%%%%%%%
\section{Spectroscopic Data} \label{sec:obs}
%%%%%%%%%%%%%%%%%%%%%%

%
% FIGURE 1
%

%%%%%%%%%%%%%%%%%%%%%%
\subsection{NOEMA}
\label{subsec:noema}
%%%%%%%%%%%%%%%%%%%%%%
The spatial resolution of SCUBA-2 is too poor to determine accurately optical/near-infrared (NIR) counterparts to submillimeter (submm) sources. Instead, interferometric observations are necessary to obtain the positions and speczs of the DSFGs. \citet{Jones21} confirmed that two DSFGs in the neighborhood of the GOODS-N (GN-CL-2 and GN-CL-3; see Table~\ref{tab:dsfgs}) have redshifts of $z = 3.133$ and $z = 3.150$, respectively, based on their CO(5--4) and CO(3--2) lines. With a target redshift of $z \sim 3.14$, we observed with NOEMA (Project Code W20DH, P.I. L. Jones) the next three brightest SCUBA-2 sources in the field (GN-CL-4, GN-CL-5, and GN-CL-6). We observed in two tunings, with Local Oscillator (LO) frequencies of 90 and 145~GHz, to target the CO(3--2) and CO(5--4) lines, respectively, in the lower sideband (LSB) of each tuning. The LO frequencies of NOEMA tunings are at the center of a 7.744~GHz gap between two 7.744~GHz sidebands, so LSBs of the band 1 and band 2 tunings cover $78.384-86.128$ GHz and $133.184-141.128$~GHz, respectively. These correspond to CO(3--2) at $z = 3.02-3.41$ and CO(5--4) at $z = 3.08-3.32$. In addition to these new data, we retrieved archival NOEMA observations of GN-CL-2 in the 3 mm band from Project S19CV (P.I.s A. Wei\ss~and S. Chapman), which cover the frequency of the expected CO(3--2) line.

All the NOEMA data were taken in the compact D-configuration to obtain total line fluxes. Beam sizes were approximately $1.9\arcsec \times3.6 \arcsec$ at 3 mm and $1.4 \arcsec \times2.0 \arcsec$ at 2 mm in the 10D configuration, and $4 \arcsec \times7 \arcsec$ at 3 mm and $2.5\arcsec \times3 \arcsec$ at 2 mm in the older 9D configuration. These data were reduced using the standard pipeline in \texttt{GILDAS} with the default data quality parameters, yielding a typical phase rms of $\sim10-30$ degrees.
A more detailed description of the calibration and imaging of these NOEMA data will be available in a companion paper (M.~Nicandro Rosenthal et al.\ 2024, in prep.), where we will discuss the physical properties of the galaxies in our sample.

We list the SCUBA-2 sources that we observed with NOEMA in Table~\ref{tab:dsfgs}.

%
% TABLE 1
%
\begin{table*}[htb]
    \begin{center}
    \caption{NOEMA Line Detections}
    \label{tab:dsfgs}
    \begin{tabularx}{\textwidth}{@{\extracolsep{6.5pt}} lccccccccc}
        \hline \hline
        Name & Hsu ID & RA & Dec & $S_{\rm 850\mu m}$ $^a$ & Project & Transition & $\nu_{\rm obs}$ & $z_{\rm spec}$ & FWHM  \\
         & & [deg] & [deg] & [mJy] & & & [GHz] & & [km s$^{-1}$] \\
        \hline
	\multicolumn{10}{c}{Protocluster Members} \\
	\hline
        GN-CL-2  & 77630 & 188.98283 & 62.37750 & $11.7\pm0.5$ & W19DG & CO(5--4) & 138.84 & 3.1504 & $405\pm26$ \\
         & & & & & S19CV & CO(3--2) & 83.30 & 3.1513 & $490\pm77$ \\
        GN-CL-3  & 85384 & 188.94433 & 62.33703 & $11.5\pm0.6$ & W19DG & CO(5--4) & 139.44 & 3.1326 & $575\pm48$ \\
         & & & & & W19DG & CO(3--2) & 83.66 & 3.1332 & $527\pm46$ \\
        GN-CL-4N & 81640 & 188.91755 & 62.35860 & $3.6\pm0.5$ & W20DH & CO(5--4) & 139.26 & $3.1381$ & $798\pm111$ \\
         & & & & & W20DH & CO(3--2) & 83.51 & $3.1410$ & $326\pm81$ \\
        GN-CL-4S & ---$^{b}$ & 188.91571 & 62.35784 & $4.5\pm0.5$ & W20DH & CO(5--4) & 139.23 & 3.1390 & $675\pm119$ \\
         & & & & & W20DH & CO(3--2) & 83.54 & 3.1392 & $537\pm193$ \\
         \hline
         \multicolumn{10}{c}{Other NOEMA Detections} \\
         \hline
        GN-CL-5  & 63299 & 188.86538 & 62.31574 & $8.6\pm0.8$ & W20DH & CO(4--3)$^{c}$ & 98.96 & $3.6588$ $^c$ & $619\pm57$ \\
        GN-CL-6  & 89441 & 188.95347 & 62.31800 & $8.2\pm0.6$ & W20DH & CO(6--5)$^c$ & 153.66 & $3.5000$ $^c$ & $843\pm130$ \\
        \hline
    \end{tabularx}
    \end{center}
    \begin{tablenotes}
        \item $a$ --- 850~$\mu$m fluxes are total fluxes from SCUBA-2 \citep{Cowie17}, except for those of GN-CL-4N and GN-CL-4S, which are from SMA (M. Nicandro Rosenthal et al.\ 2024, in prep.). The total flux of GN-CL-4 from SCUBA-2 is $10.8 \pm 0.5$ mJy.
        \item $b$ --- GN-CL-4S is undetected in ground-based and HST optical-to-NIR imaging (M.~Nicandro Rosenthal et al.\ 2024, in prep.).
        \item $c$ --- Only one emission line is detected. The stated redshifts assume the detected line is a CO emission line and use the solution closest to the photzs from \citet{Hsu19}.
    \end{tablenotes}
\end{table*}

%%%%%%%%%%%%%%%%%%%%%%
\subsection{Keck Multi-Object Spectroscopy}
\label{subsec:keck}
%%%%%%%%%%%%%%%%%%%%%%

%
% TABLE 2
%
\begin{table*}
	\centering
	\caption{Keck Targets and Detection Fractions}
	\begin{tabularx}{\linewidth}{l  @{\extracolsep{16pt}}  ccc  @{\extracolsep{15pt}}  ccccc}
	\hline \hline
	& \multicolumn{3}{c}{LRIS} & \multicolumn{5}{c}{MOSFIRE} \\
	\cmidrule{2-4} \cmidrule{5-9}
	(1) & (2) & (3) & (4) & (5) & (6) & (7) & (8) & (9) \\
	Type & $N_{\rm obs}$ & $N_{\rm det}$ & $f_{\rm det}$ & $N_{\rm obs}$ & $N_{\rm det}$ & $f_{\rm det}$ & $f_{\rm det}^{(K\leq23)}$ & $f_{\rm det}^{(K>23)}$ \\
	\hline
        smm & 4 & 2 & 50.0\% & 23 & 6 & 26.1\% & $4/11 = 36.4\%$ & $2/12 = 16.7\%$ \\
        z3a  & 45 & 9 & 20.0\% & 141 & 67 & 47.5\% & $15/28 = 53.6\%$ & $52/113 = 46.0\%$ \\
        z3b  & 25 & 5 & 20.0\% & 140 & 37 & 26.4\% & $5/20 = 25.0\%$ & $32/120 = 26.7\%$ \\
        Total main targets & 74 & 16 & 21.6\% & 304 & 110 & 36.2\% & $24/59=40.7\%$ & $86/245=35.1\%$ \\
        \hline
        cont & 24 & 3 & 12.5\% & --- & --- & --- & --- & --- \\
        20cm & 7 & 0 & 0.0\% & --- & --- & --- & --- & --- \\
        hiz & 30 & 2 & 6.67\% & --- & --- & --- & --- & --- \\
        z2 & --- & --- & --- & 85 & 23 & 27.1\% & $10/28 = 35.7\%$ & $13/57 = 22.8\%$ \\
        Total filler targets & 62 & 5 & 8.1\% & 85 & 23 & 27.1\% & $10/28 = 35.7\%$ & $13/57 = 22.8\%$ \\
        \hline
        Grand total & 136 & 21 & 15.4\% & 389 & 133 & 34.2\% & $34/87 = 39.0\%$ & $99/302 = 32.8\%$ \\
	\hline
	\end{tabularx}
	\label{tab:keck_desc}
\end{table*}

In addition to massive and rare galaxies like DSFGs, protoclusters have many tens of less massive members, i.e., ``normal'' star-forming galaxies, which can be detected with optical and/or NIR spectroscopy. There have been many spectroscopic surveys of the GOODS-N \citep[e.g.,][]{Barger08, Cooper11, Brammer12, Kriek15}, but these surveys targeted the original GOODS \citep{Giavalisco04} or CANDELS \citep{Grogin11} HST fields (cyan outline in Figure \ref{fig:map}). They do not cover the region around the $z \sim 3.14$ DSFGs discovered by \citet{Jones21} (red squares in Figure \ref{fig:map}). 

We therefore conducted a new Keck spectroscopic survey northwest of the original GOODS-N field. Using the ground-based and Spitzer/IRAC catalog of \citet{Hsu19}, we targeted optical/NIR-selected galaxies with photometric redshifts (photzs) $z_{\rm phot} \sim 3$. This region covers the overdensity of $z_{\rm phot} \sim 3.1$ Lyman Break Galaxies (LBGs) identified by \citet{Jones21}. We prioritized optical/NIR sources within $5''$ of the centroid of an 850\,$\mu$m source (``smm'' in Table \ref{tab:keck_desc}), which might be counterparts to the 850\,$\mu$m source, then sources with $3.0 \leq z_{\rm phot} \leq 3.3$ (``z3a''), and finally sources with $2.7 \leq z_{\rm phot} < 3.0$ or $3.3 < z_{\rm phot} \leq 3.6$ (``z3b''). In addition to these main science targets, we included filler objects to maximize the number of slits per mask (see Sections~\ref{sec:LRIS} and \ref{sec:MOSFIRE}). In Table~\ref{tab:keck_desc}, we give the number of galaxies of each type that we observed and detected with LRIS and MOSFIRE.

%%%%%%%%%%%%%%%%%%%
\subsubsection{LRIS} \label{sec:LRIS}
%%%%%%%%%%%%%%%%%%%
We conducted one night of observations with the Low-Resolution Imaging Spectrometer \citep[LRIS;][]{LRIS} on March 27, 2022. For these observations we used the 5600\,\AA~dichroic, with the 600/4000 grism on the blue side of LRIS to target Ly$\alpha$ emission at $\lambda_{\rm obs} \approx 5000$\,\AA,~and the 400/8500 grating on the red side with a central wavelength of 8500\,\AA~to target UV absorption features. We imposed a magnitude limit of $V \leq 25.2$ for our LRIS targets. We observed each slitmask for $3 \times 1800$\,s on the blue side and $6 \times 900$\,s on the red side, for a total of 90 minutes per mask (the red side requires shorter exposures to minimize the impact of cosmic rays on its thicker CCD). We reduced these data with an in-house IDL package to apply all calibrations, including sky subtraction, flat fielding, bias correction, and wavelength calibration. For details, we refer the reader to Section~2 of \citet{Cowie96} and Section~2 of \citet{Cowie04}.

For LRIS filler objects, we targeted radio sources detected in the 1.4 GHz Karl G. Jansky Very Large Array (VLA) mosaics from \citet{Morrison10} and \citet{Owen18} (``20cm''), and candidate high-redshift sources with $z_{\rm phot} > 2.5$ (``hiz''). In each mask, we also included several bright $V$-band continuum objects (``cont'') to aid in our reductions. 

%%%%%%%%%%%%%%%%%%%%%%%%
\subsubsection{MOSFIRE} \label{sec:MOSFIRE}
%%%%%%%%%%%%%%%%%%%%%%%%
We also observed with the Multi-Object Spectrometer For Infra-Red Exploration \citep[MOSFIRE;][]{MOSFIRE1, MOSFIRE2} over a total of four nights, on March 17-18, 2021 and March 25-26, 2022. We used $K$-band multi-object spectroscopy to target the [O\,{\sc iii}]\,$\lambda\lambda$4959,5007\,\AA~doublet at $3.03 \leq z \leq 3.59$ ($\lambda_{\rm obs} = 2.0-2.3~\mu$m). We imposed a magnitude limit of $K_s \leq 24.5$ on the large majority of our MOSFIRE targets, with four exceptions, which we included when additional $K_s \leq 24.5$ objects could not be fit on a mask. Of these, only one was detected, at $z = 3.3198$. This redshift is sufficiently high that it does not affect our overdensity analysis in Section \ref{sec:protoclusters}.

We observed each slitmask for 5 repeats of an ABBA dither pattern with 180~s exposures, giving a total of 1~hr of exposure time per mask. As with the LRIS data, we reduced these MOSFIRE data using our own IDL package, the details of which are discussed in Section~2.4 of \citet{Cowie16}.

For MOSFIRE filler objects, we observed sources with $1.9 < z_{\rm phot} < 2.7$ (``z2''), targeting the [N\,{\sc ii}]\,$\lambda\lambda$6549,6585\,\AA~+ H$\alpha$ complex in the $K$-band at $2.04 \leq z \leq 2.50$.

%%%%%%%%%%%%%%%%
\subsubsection{Completeness}
%%%%%%%%%%%%%%%%
As a rough estimate of our completeness, we consider how many galaxies that satisfy our ``z3a'' and ``z3b'' criteria within a circle of radius 10 cMpc around the average position of GN-CL-2 and GN-CL-3 we observed. We selected this radius because simulated and observed protoclusters have been shown to contain most of their galaxies and mass within $\sim$10 cMpc \citep[e.g.,][]{Chiang13, Casey16, Alberts22}. This corresponds to an angular radius of $5.4'$ at $z = 3.1$ (yellow circle in Figure \ref{fig:map}). Within this circle, we observed 100/127 (78.7\%) of ``z3a'' and 142/258 (55.0\%) of ``z3b'' galaxies that satisfied either our $K_s \leq 24.5$ or $V \leq 25.2$ magnitude limit.

%%%%%%%%%
\section{Results}
\label{sec:results}
%%%%%%%%%

%
% FIGURE 2
%
\begin{figure}[htb]
    \centering
    \includegraphics[width=\linewidth]{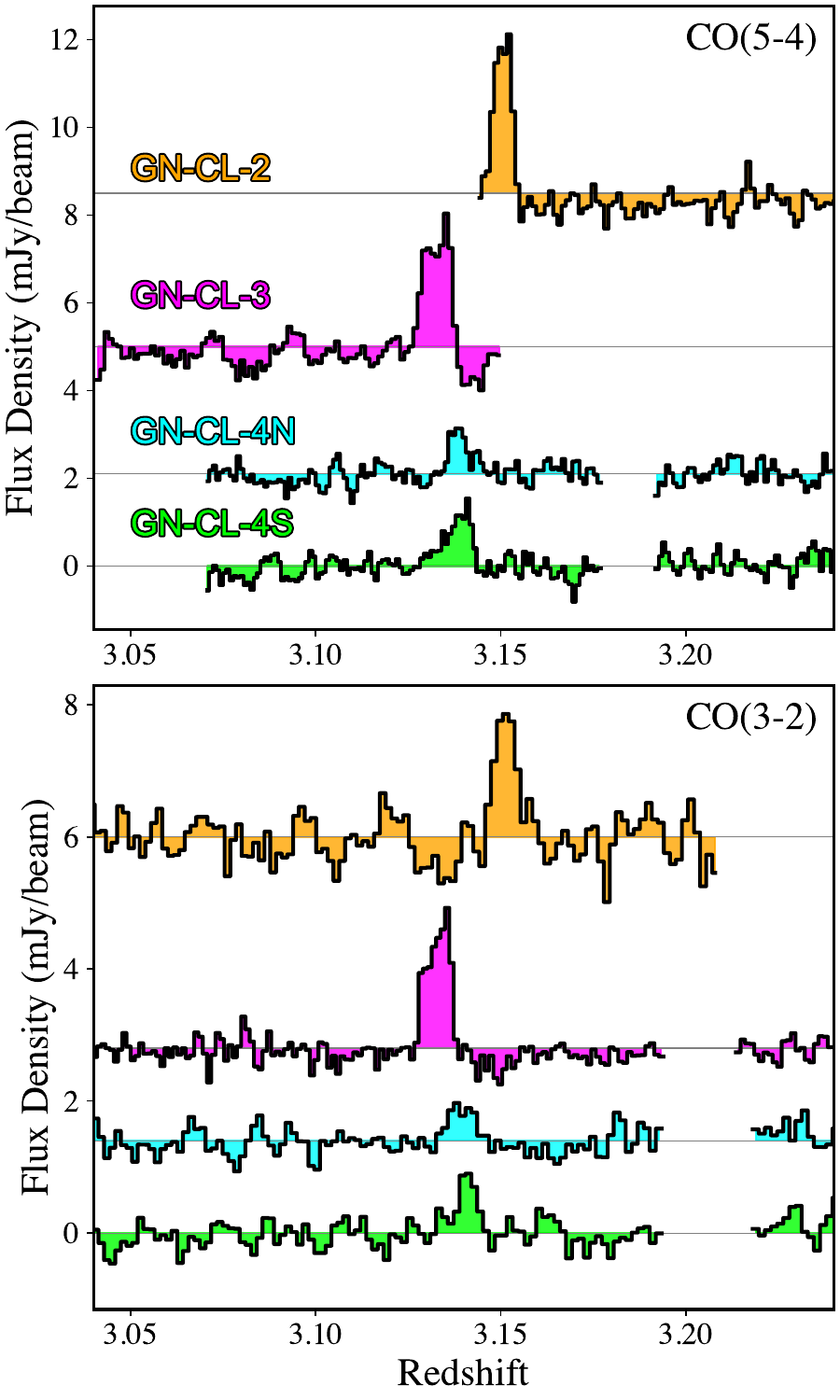}
    \caption{NOEMA CO(5--4) and CO(3--2) spectra of the four DSFGs at $z = 3.143 \pm 0.01$. From top to bottom the spectra are for GN-CL-2, GN-CL-3, GN-CL-4N, and GN-CL-4S. The spectra have been Gaussian smoothed and arbitrary flux offsets were added for display purposes.
    }
    \label{fig:noema_spec}
\end{figure}

%%%%%%%%%%%%%%%%%%%%%%%%%%%%%%
\subsection{NOEMA CO Detections} \label{sec:noema_det}
%%%%%%%%%%%%%%%%%%%%%%%%%%%%%%
We detected at least one strong CO line in each of the three SCUBA-2 sources we observed as part of NOEMA Project W20DH. As described in the following paragraphs, these detections were sufficient to confirm or rule out the galaxies' protocluster membership. Following the procedure in \citet{Jones21}, we generated a one-dimensional (1D) spectrum from the brightest continuum spaxel in the NOEMA image cube for each sideband, and we inspected these by hand for possible emission lines. In Figure \ref{fig:noema_spec}, we show the CO(5--4) and CO(3--2) spectra of the four DSFGs with line emission detected in this way (e.g., GN-CL-2, GN-CL-3, GN-CL-4N, and GN-CL-4S). To measure line widths and redshifts, we fit single Gaussians to the 1D spectra. We give the results of this fitting procedure in Table~\ref{tab:dsfgs}, and we describe the individual sources below:

\paragraph{GN-CL-2 (archival)} This source was already found to have a 4.3 mJy line centered at 138.9 GHz by \citet{Jones21}. Those authors used the March 2021 MOSFIRE observations presented in this paper to confirm that this source had $z_{\rm spec} = 3.150$, which shows that this line is CO(5--4). 
We inspected the archival 3\,mm observations of GN-CL-2 and measured a CO(3--2) line at the same redshift.

\paragraph{GN-CL-3 (archival)} Like GN-CL-2, GN-CL-3 was originally discussed in \citet{Jones21}, but we summarize their results here for completeness. \citet{Jones21} identified two bright lines in the NOEMA spectra of GN-CL-3, centered at 139.5 and 83.7 GHz. These lines are consistent with CO(5--4) and CO(3--2), respectively, at a redshift of $z = 3.133$.

\paragraph{GN-CL-4} This source is resolved by NOEMA into two component galaxies separated by $7''$. In both of these components, we detected strong line emission in the 2\,mm and 3\,mm NOEMA spectra. Based on their sky frequencies, we find that these lines are CO(5--4) and CO(3--2) at $z_{\rm spec} = 3.139$.

\paragraph{GN-CL-5} We measure a single strong line in GN-CL-5 at 99.0 GHz, which is consistent with either CO(2--1) at $z = 1.330$ or CO(4--3) at $z = 3.659$. All other CO lines at $1 \leq z \leq 8$ are ruled out by the non-detection of additional lines in our spectral coverage. The NOEMA source is coincident with a galaxy with $z_{\rm phot} = 3.33 \pm 1.12$, favoring CO(4--3) at $z = 3.659$. 

\paragraph{GN-CL-6} We detect one strong line in GN-CL-6 at 153.7 GHz, which is inconsistent with $z \sim 3.1$. Of the many possible redshift solutions for GN-CL-6, two are ruled out: CO(7--6) at $z = 4.240$ is ruled out by the lack of a [C\,\textsc{i}]($^3$P$_2$--$^3$P$_1$) line, and CO(8--7) is ruled out by the lack of additional CO lines at lower frequencies. The line emission is coincident with a $z_{\rm phot} = 4.300$ galaxy from \citet{Hsu19}. GN-CL-6 has a 450\,$\mu$m/850\,$\mu$m flux ratio of $\approx 2.2$, whereas most $z \sim 1$ DSFGs have ratios of $\geq 3$ \citep[][]{Barger22, Cowie22}, suggesting that it lies at $z \gtrsim 2$. We therefore find the line is most likely CO(6--5) at $z = 3.5000$.

%%%%%%%%%%%%%%%%%%%%%%
\subsection{Keck Spectroscopic Detections}
%%%%%%%%%%%%%%%%%%%%%%
We detected a total of 24/138 of our LRIS targets (17.4\%) and 132/385 of our MOSFIRE targets (34.3\%), including 2 galaxies that we detected with both instruments, one of which is GN-CL-2. This yields a total of 154 new speczs from Keck. In Table \ref{tab:redshifts}, we give their positions, magnitudes, and redshifts.

\begin{figure}
	\centering
	\includegraphics[width=\linewidth]{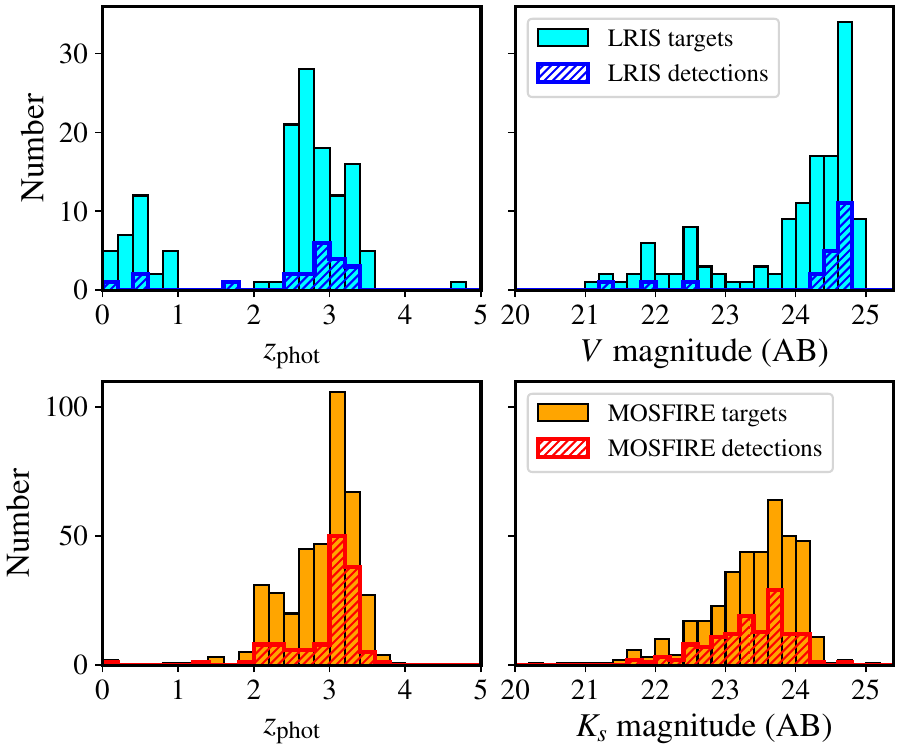}
	\caption{Histograms of the photzs and broadband magnitudes of our targets (filled bars) and detections (hatched bars). \textit{Top}: LRIS observations and detections, which were selected with a limiting magnitude of $V \leq 25.2$. \textit{Bottom}: MOSFIRE observations and detections, which were selected with a limiting magnitude of $K_s \leq 24.5$. The bottom-left panel shows a higher MOSFIRE detection fraction for galaxies with $z_{\rm phot} \sim 3$, which reflects the detectability of the [O{\sc iii}] and H$\beta$ complex in the $K$-band at these redshifts.}
	 \label{fig:detection_hist}
\end{figure}

As we show in Figure \ref{fig:detection_hist}, we obtained the highest detection rate in galaxies with $z_{\rm phot} \sim 3.1$, where the [O{\sc iii}]+H$\beta$ complex is detectable in $K$-band spectroscopy. We find a total of 78 galaxies have $3.0 \leq z_{\rm spec} \leq 3.3$. Overall, 89.2\% of our detections have normalized absolute redshift deviations $|z_{\rm spec} - z_{\rm phot}| / (1 + z_{\rm spec}) \leq 0.15$, with a normalized median absolute deviation $\sigma_{\rm NMAD} = 0.042$, which is consistent with expectation for $z > 1$ sources from \citet{Hsu19}. We give the number of observed and detected galaxies for each selection type in Table \ref{tab:keck_desc}.

As listed in Table \ref{tab:keck_desc}, we had 69 ``z3a'' targets that we detected with either LRIS or MOSFIRE. Of these, 47 (68.1\%) had $3.0 \leq z_{\rm spec} \leq 3.3$. We also had 44 ``z3b'' detections, of which 23 (52.3\%) had $3.0 \leq z_{\rm spec} \leq 3.3$ and a further 11 (25.0\%) had other redshifts between $2.7 \leq z_{\rm spec} \leq 3.6$. In addition to the NOEMA linescans of the brightest SCUBA-2 sources, we also detected a total of 10 ``smm'' targets: 2/4 targeted with LRIS and 9/26 targeted with MOSFIRE. Both of these counts include GN-CL-2, which was detected with all three instruments (including NOEMA). %at $z=3.152\pm0.002$; and 
GN-CL-3 was also detected with MOSFIRE. Apart from the NOEMA-detected DSFGs detailed in Section \ref{sec:noema_det}, none of the ``smm'' detections have speczs similar to those of the other DSFGs. Thus, while we give their positions and redshifts in Table \ref{tab:redshifts}, we do not include them in our analysis for the remainder of this work.

%
% FIGURE 3
%

%%%%%%%%%%%%%%
\subsection{Redshift Spikes}
\label{sec:spikes}
%%%%%%%%%%%%%%

%
% TABLE 3
%
\begin{table*}
	\begin{center}
	\caption{Spectroscopic Detections from our NOEMA and Keck Surveys}
	\label{tab:redshifts}
	\begin{tabularx}{\linewidth}{@{\extracolsep{13pt}} ccccccccccc}
        \hline \hline
        (1) & (2) & (3) & (4) & (5) & (6) & (7) & (8) & (9) & (10) & (11) \\
        Type & Hsu ID & RA & Dec. & $V$ & $R$ & $K_s$ & $z_{\rm phot}$ & $z_{\rm opt}$ $^{a}$ & $z_{\rm NIR}$ $^{b}$ & $z_{\rm CO}$ $^{c}$ \\
         &  & [deg] & [deg] & [AB] & [AB] & [AB] &  &  &  & \\
        \hline
	smm & --- & 188.91560 & 62.3569 & --- & --- & --- & --- & $-3.0$ & $-3.0$ & 3.139 \\
	z3a & 80570 & 188.89441 & 62.3621 & 25.27 & 25.18 & 23.94 & 3.1831 & $-3.0$ & 3.1438 & $-3.0$ \\
	z3a & 79176 & 188.90288 & 62.3703 & 25.22 & 24.76 & 24.17 & 3.1209 & $-3.0$ & 3.1344 & $-3.0$ \\
	z3b & 79100 & 188.93536 & 62.3724 & 24.94 & 24.73 & 23.20 & 3.4462 & $-3.0$ & 3.1388 & $-3.0$ \\
	z3b & 78192 & 188.79875 & 62.3750 & 26.79 & 25.67 & 21.98 & 3.3747 & $-3.0$ & 3.1485 & $-3.0$ \\
	$\cdots$ & $\cdots$ & $\cdots$ & $\cdots$ & $\cdots$ & $\cdots$ & $\cdots$ & $\cdots$ & $\cdots$ & $\cdots$ & $\cdots$ \\
	z3b & 76939 & 188.97598 & 62.3792 & 24.99 & 24.67 & 24.09 & 3.3360 & $-3.0$ & 3.1428 & $-3.0$ \\
	z3a & 76454 & 188.97029 & 62.3839 & 24.67 & 24.76 & 23.02 & 3.1756 & $-3.0$ & 3.1431 & $-3.0$ \\
	smm & 81640 & 188.91751 & 62.3586 & 28.38 & 26.45 & 22.47 & 3.4000 & $-3.0$ & $-1.0$ & 3.139 \\
	z3b & 76319 & 188.97017 & 62.3744 & 24.31 & 23.49 & 22.26 & 3.3299 & $-3.0$ & 3.1429 & $-3.0$ \\
	z3a & 68202 & 189.19633 & 62.4396 & 25.54 & 25.52 & 23.64 & 3.0964 & $-3.0$ & 3.151 & $-3.0$ \\
	\hline
	\end{tabularx}
	\end{center}
	\begin{tablenotes}
	\item Columns are (1) Object type (Section \ref{subsec:keck}); (2)--(4) sequence number, right ascension, and declination from \citet{Hsu19}, where available; otherwise, the positions are from NOEMA (this work); (5)--(8) $V$-band, $R$-band, and $K_s$-band magnitude and photz from \citet{Hsu19}; (9)--(11) specz from LRIS, MOSFIRE, and NOEMA. For all speczs, positive values are measurements, $-1.0$ are non-detections with that instrument, and $-3.0$ were not observed with that instrument.
	    	\item The complete table is available in machine readable form in the online version.
	\end{tablenotes}
\end{table*}

%
% FIGURE 4
%
\begin{figure}[t]
    \centering
    \includegraphics[width=\linewidth]{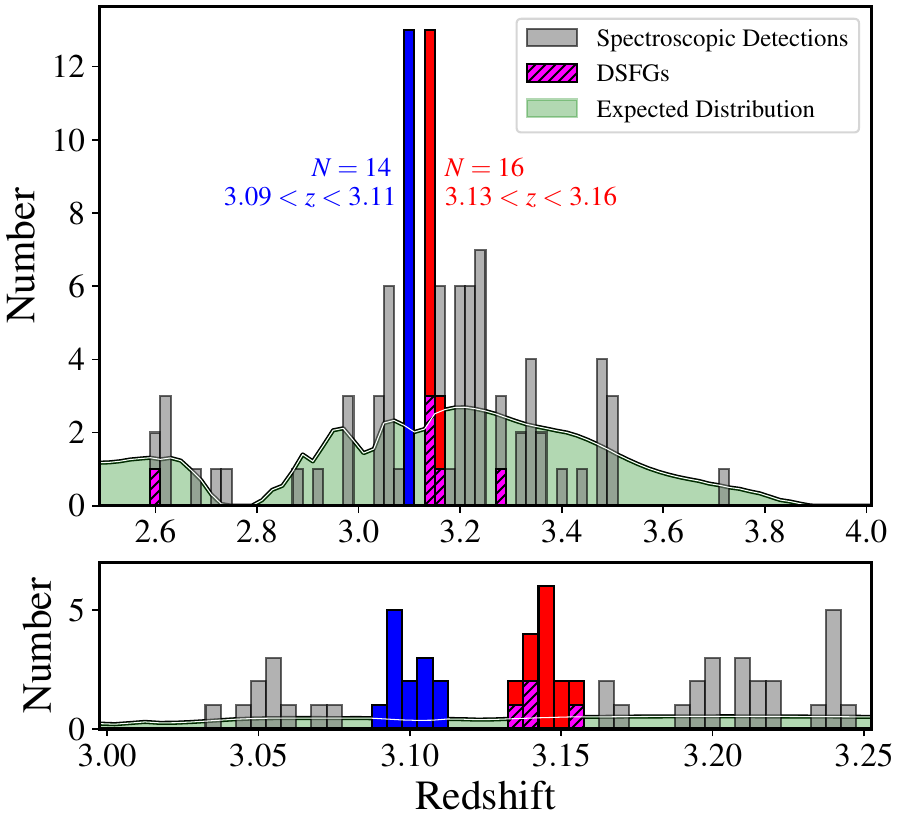}
    \caption{
    \textit{Top:} Histogram of speczs from our combined NOEMA and Keck surveys. The green region in the background shows the expected number of galaxies per $\Delta z = 0.02$ bin based on our photz selection (see Appendix \ref{sec:photoz_pdf}). The red and blue colored bars indicate GNCL-z3.14 and GNCL-z3.10, respectively (see Section \ref{sec:protoclusters}). Hatched magneta bars indicate DSFGs with speczs from our survey. \textit{Bottom:} The same histogram, but with bins of $\Delta z = 0.005$ and zoomed in.}
    \label{fig:z_hist}
\end{figure}

We show the histogram of all speczs in our survey in the top panel of Figure \ref{fig:z_hist} with bins of $\Delta z = 0.02$, which corresponds to a co-moving depth of $d_{\rm c} \sim 18\,{\rm cMpc}$ at $z \sim 3$. This binning was chosen such that the co-moving depth is similar to the size of our survey area (the 10 cMpc radius circle in Figure \ref{fig:map}). The redshift histogram includes two ``spikes'', reminiscent of previous spectroscopic confirmations of protoclusters \citep[e.g.,][]{Steidel98, Chapman09}, which contain 16 and 14 galaxies at $z = 3.13-3.16$ and $z = 3.09-3.11$, respectively. Henceforth, we will refer to these as GNCL-z3.14 and GNCL-z3.10, respectively. We note that while the bin width here does maximimze the contrast of the spikes, we do not calculate the galaxy overdensity directly from these spikes (we describe our actual methodology in Section \ref{sec:protoclusters}). In the bottom panel of Figure \ref{fig:z_hist}, we show a zoom-in of the histogram with a binning of $\Delta z = 0.005$ ($\Delta d_c \sim 4.5\,{\rm cMpc}$).

The higher-redshift spike, GNCL-z3.14, contains 4 DSFGs, which are denoted by the hatched magenta bars in Figure \ref{fig:z_hist}. Meanwhile, GNCL-z3.10 has no confirmed DSFGs. The number of galaxies per bin and the volume corresponding to each bin are similar to the $z = 3.09$ SSA22 protocluster from \citet{Steidel98}, but we detect two overdensities in close proximity.

%
% FIGURE 5
%
\begin{figure*}[htb]
    \centering
    \includegraphics[width=\linewidth]{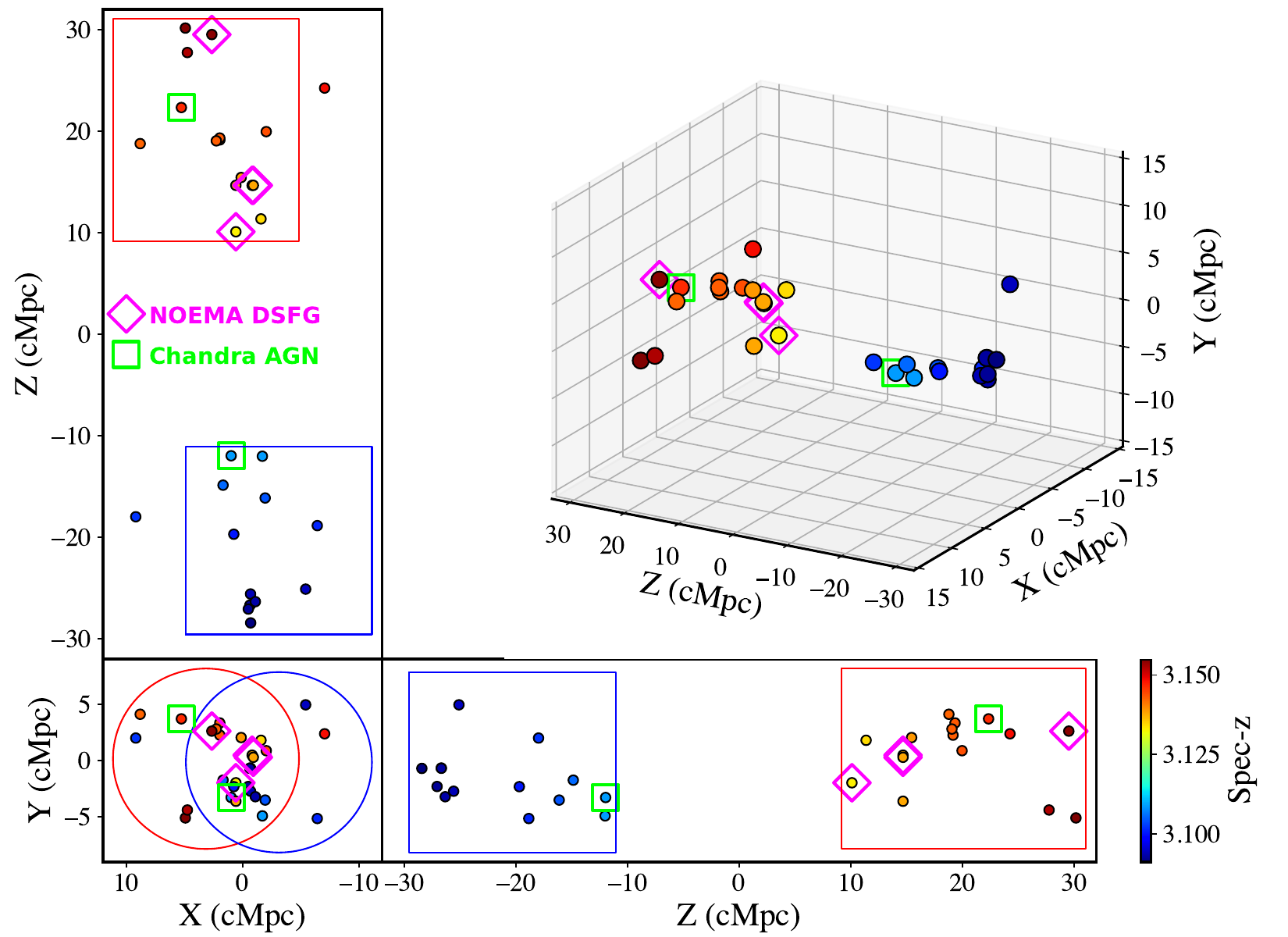}
    \caption{\textit{Top right:} A 3D map of the protoclusters. \textit{Left, bottom-left, and bottom:} 2D projections of the same region. In all panels, the radial offsets (Z) in physical units are calculated assuming no peculiar velocities. Magenta diamonds indicate NOEMA-detected DSFGs, and green squares indicate \textit{Chandra}-detected AGNs. The red and blue outlines are the projected boundaries of the volumes used to calculate the overdensities of GNCL-z3.14 and GNCL-z3.10, respectively.}
    \label{fig:3dmap}
\end{figure*}

%%%%%%%%%%%%%%%%%%
\section{Two Galaxy Overdensities}
\label{sec:protoclusters}
%%%%%%%%%%%%%%%%%%
Our spectroscopic surveys show two clear overdensities of galaxies at $z = 3.10$ and $z = 3.14$. However, not all high-redshift overdensities eventually collapse into clusters. In this work, we take ``protocluster'' to mean objects that will  collapse into clusters with $M_{z=0} \geq 10^{14}~{\rm M}_\sun$ \citep[][]{Chiang13, Overzier16, Alberts22}. In this section, we calculate the precise galaxy overdensities of these spikes and convert them into matter overdensities and total masses. Galaxy and mass overdensities are defined, respectively, as
\begin{equation}
    \delta_{\rm gal,obs} \equiv \frac{N_{\rm obs} - N_{\rm exp}}{N_{\rm exp}} \label{eqn:d_gal} \,,
\end{equation}
and 
\begin{equation}
    \delta_{\rm m} \equiv \frac{\rho_{\rm m} - \bar{\rho}_{\rm m}(z)}{\bar{\rho}_{\rm m}(z) }  \,,
    \label{eqn:d_m}
\end{equation}
where $N_{\rm obs}$ is the number of galaxies observed in a given volume, $N_{\rm exp}$ is the number of galaxies expected in that same volume, $\rho_{\rm m}$ is the observed total matter density, and $\bar{\rho}_{\rm m}(z)$ is the average matter density at a given redshift. Following \citet{Steidel98}, galaxy and matter overdensities are related by
\begin{equation}
	1 + b\,\delta_{\rm m} = C (1 + \delta_{\rm gal,obs}) \label{eqn:dgal_dm} \,.
\end{equation}
Here $C \equiv V_{\rm true} / V_{\rm app}$ is the redshift distortion factor that relates the true volume occupied by the galaxies to the apparent volume based on their minimum and maximum observed redshifts, and $b$ is the linear galaxy bias, which is a dimensionless scalar that depends on the properties of the observed galaxies, as well as the scale on which the overdensity is measured \citep{Chiang13}. Collapsing structures have $C < 1$ and virialized structures have $C > 1$. The subscript ``obs'' refers to the fact that $\delta_{\rm gal,obs}$ is measured for an apparent volume without correcting for the redshift distortion. From \citet{Lahav91}, the redshift distortion factor and mass overdensity are related by
\begin{equation}
	C = 1 + f - f (1 + \delta_{\rm m})^{1/3} \label{eqn:C} \,,
\end{equation}
where $f = \Omega_{\rm m}(z)^{4/7}$. For $z \sim 3.1$, $f \sim 0.98$. By simultaneously solving Equations \ref{eqn:dgal_dm} and \ref{eqn:C} with fixed values of $b$ and $\delta_{\rm gal,obs}$, it is straightforward to solve for $C$ and $\delta_{\rm m}$.

For each of the spikes described in Section \ref{sec:results}, we measure the overdensity in a cylindrical volume. The depth of the cylinder is defined by the lowest and highest redshift galaxies in each group, and we take the radius of the cylinder to be 8 cMpc ($4.3'$ at $z = 3.1$), which is the typical effective radius of the most massive $z\sim3$ protoclusters \citep[e.g.,][]{Chiang13}. Using this definition, we obtain apparent volumes of $V_{\rm app}^{\rm (z3.14)} = 4030~{\rm cMpc}^3$ and $V_{\rm app}^{\rm (z3.10)} = 3310~{\rm cMpc}^3$. We show the boundaries of volumes A and B in red and blue, respectively, in Figure \ref{fig:3dmap}.
Measuring $N_{\rm obs}$ is then as simple as counting the number of spectroscopically confirmed galaxies within each of these volumes. To calculate $N_{\rm exp}$, we use two methods: one by photz selection, and one by comparing to broadband luminosity functions at $z \sim 3$.

%%%%%%%%%%%%%%%%%%%%%%%%%
\subsection{Comparison to Photometric Redshifts}
\label{sec:photoz_overdensity}
%%%%%%%%%%%%%%%%%%%%%%%%%
To estimate the galaxy overdensities of the two redshift spikes, we take advantage of the fact that all but two\footnote{GN-CL-1 was not included due to its proximity to a bright star, and GN-CL-4S is optically dark and therefore does not have a \citet{Hsu19} ID.} of our targets have \texttt{eazy} \citep{Brammer08} photzs from \citet{Hsu19}. \citet{Hsu19} tested their photz fits against a subset of their sample that already had good speczs. If we assume that our targets follow the same distribution in  $(z_{\rm spec} - z_{\rm phot})/ (1 + z_{\rm phot})$ as these, then for any given input photz, we can calculate a probability distribution function (PDF) of the true redshift. We take $P(z_{\rm min} \leq z \leq z_{\rm max})$ to be the integral of this PDF over some redshift bounds. We take the probability of detecting each galaxy to be the product of this probability and the estimated detection fraction for galaxies in a given photz range. We describe these calculations in detail in Appendix \ref{sec:photoz_pdf}. 

Using this method, we expect a total of $N_{\rm exp} = 1.97$ galaxies in GNCL-z3.14 and 1.49 galaxies in GNCL-z3.10. From Equation \ref{eqn:d_gal}, this gives galaxy overdensities of $\delta_{\rm gal,obs}^{\rm (z3.14)} = 6.10$ and $\delta_{\rm gal,obs}^{\rm (z3.10)} = 7.07$. These overdensity estimates are not corrected for the fact that there were a greater than expected number of $z_{\rm phot} \sim 3.1$ galaxies in our survey area to begin with, as first identified by \citet{Jones21}, and may therefore be considered lower limits.

%%%%%%%%%%%%%%%%%%%%%%%%%%%%%%%
\subsection{Comparison to Broadband Luminosity Functions}
\label{sec:lf_overdensity}
%%%%%%%%%%%%%%%%%%%%%%%%%%%%%%%

%
% FIGURE 6
%
\begin{figure}
	\centering
	\includegraphics[width=\linewidth]{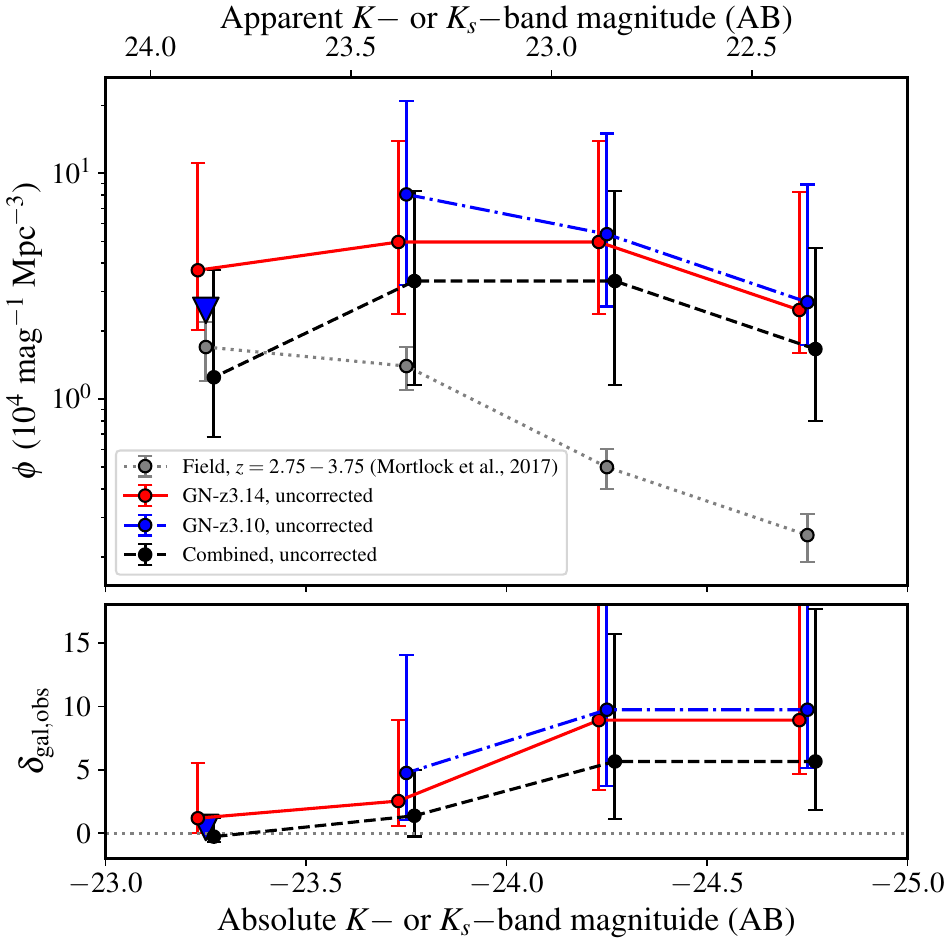}
	\caption{
	\textit{Top:} $K_s$-band LFs of GNCL-z3.10 (blue), GNCL-z3.14 (red), and the combined potential hyper-protocluster (black; see Section \ref{sec:hyper}) compared with the $z = 2.75-3.75$ field $K$-band LF of \citet{Mortlock17} (gray). The LFs for the protoclusters in this work are not corrected for completeness; thus, the actual number densities within the protoclusters may be higher. \textit{Bottom:} Overdensities of the protoclusters relative to the \citet{Mortlock17} LF in each bin. The brighter bins are likely more accurate, as our detection fraction and completeness for the fainter bins are substantially lower.}
	\label{fig:klf}
\end{figure}

To obtain the most direct estimate of $N_{\rm exp}$, we take the number density of field galaxies selected similarly to those in our spectroscopic survey and multiply by the observed volume. We use the $K$-band luminosity function (LF) of \citet{Mortlock17}, since we observed the majority of our targets with MOSFIRE using a $K$-band magnitude cut. In Figure \ref{fig:klf}, we show that in all bins of $K \leq 24.5$, we detect more galaxies than would be expected for the co-eval field. To estimate the total value of $N_{\rm exp}$, we summed all the bins with $K \leq 24.5$~mag multiplied by the width of the bins, i.e., 0.5 mag, and by the overall detection fraction of our MOSFIRE survey. For $2.75 < z < 3.75$, \citet{Mortlock17} find a typical source density of $n_{\rm gal} = 3.53~{\rm cMpc}^{-3}$ for galaxies with $M_K \leq -22.5$ ($K \leq 24.62$ at $z = 3.1$). Based on this, we find $N_{\rm exp}^{\rm (z3.14)} = 1.29$ and $N_{\rm exp}^{\rm (z3.10)} = 1.17$, which correspond to observed galaxy overdensities of $\delta_{\rm gal,obs}^{\rm (z3.14)} = 9.10$ and $\delta_{\rm gal,obs}^{\rm (z3.10)} = 9.28$. Note that for these calculations, we use $N_{\rm obs}^{\rm (z3.14)} = 13$ rather than 14, as GN-CL-4S is not detected in the $K$-band, and hence it must be excluded from the comparison to the $K$-band LF.

%%%%%%%%%%%%%%%%%%%%%%%%%%
\subsection{Mass Overdensities and Total Masses}
\label{sec:masses}
%%%%%%%%%%%%%%%%%%%%%%%%%%

Having calculated $\delta_{\rm gal,obs}$, we can now obtain the matter overdensity $\delta_{\rm m}$ for each protocluster by simultaneously solving Equations \ref{eqn:dgal_dm} and \ref{eqn:C}. We assume a bias of $b=2.71$, which is appropriate for $M_\star > 10^{10}~{\rm M}_\sun$ galaxies at $z\sim3$ measured over a $15~{\rm cMpc}^3$ volume \citep{Chiang13}, since $V_{\rm app}^{\rm (z3.14)} \sim V_{\rm app}^{\rm (z3.10)} \sim (15~{\rm cMpc})^3 = 3375~{\rm cMpc}^3$. We measure redshift distortions of $C \approx 0.6-0.7$ for both GNCL-z3.14 and GNCL-z3.10 using both of our overdensity calculation methods, finding slightly higher distortions for the photoz-based estimates. We also measure matter overdensities of $\delta_{\rm m} > 1$ for both of these methods, finding slightly higher matter overdensities for the LF-based estimates. We list the specific values in Table \ref{tab:protocluster_properties}. Following \citet{Steidel98}, we convert these values into a redshift zero total mass using the formula
\begin{equation}
	M_{z=0} = C\,V_{\rm app}  \, \bar{\rho}_{\rm m} \, (1 + \delta_{\rm m}) \,,
\end{equation}
where $\bar{\rho}_{\rm m} = 4.08 \times 10^{10}~{\rm M}_\sun~{\rm cMpc}^{-3}$ at $z = 3.10$. This gives $M_{z=0} \simeq (6-7) \times 10^{14}~{\rm M}_\sun$ for both protoclusters using both methods. Again, we list the specific masses in Table~\ref{tab:protocluster_properties}. We note, however, that these overdensities, and thus mass calculations, are subject to uncertainties, both in the number of expected galaxies and in the Poisson uncertainty in our number counts. We describe these uncertainties, and the resulting uncertainties in the final masses, in Section \ref{sec:monte_carlo}.

%
% TABLE 4
%
\begin{table}
	\begin{center}
	\caption{Measured Properties of the $z\sim3.1$ Protoclusters}
	\label{tab:protocluster_properties}
	\begin{tabularx}{\linewidth}{lccc}
		\hline \hline
		 & GNCL-z3.14 & GNCL-z3.10 & Combined \\
		\hline
		$z_{\rm min}$ & 3.133 & 3.090 & 3.090 \\
		$z_{\rm max}$ & 3.155 & 3.110 & 3.155 \\
		RA [deg.] & 188.99307 & 188.87337 & 188.93321 \\
		Dec. [deg.] & 62.35593 & 62.35292 & 62.35444 \\
		$N_{\rm specz}$ & 14 & 12 & 25 \\
		$V_{\rm app}$ [cMpc$^3$] & 4030 & 3310 & 11\,800 \\
		\hline
		\multicolumn{4}{c}{Photometric Redshift Method (Section \ref{sec:photoz_overdensity})} \\
		\hline
		$N_{\rm exp}$ & 1.97 & 1.49 & 5.69 \\
		$\delta_{\rm gal,obs}$ & 6.10 & 7.07 & 3.39 \\
		$C$ & 0.670 & 0.642 & 0.770 \\
		$V_{\rm true}$ [cMpc$^3$] & 6570 & 5800 & 15\,600 \\
		$\delta_{\rm m}$ & 1.39 & 1.54 & 0.88 \\
		$M_{z=0}~[10^{14}\,{\rm M}_\sun]$ & 6.40 & 6.02 & 12.0 \\
		$\delta_{\rm L}(z=0)$ & 2.24 & 2.34 & 1.73 \\
		$z_c$ & 0.53 & 0.61 & 0.05 \\
		\hline
		\multicolumn{4}{c}{Luminosity Function Method (Section \ref{sec:lf_overdensity})} \\
		\hline
		$N_{\rm exp}$ & 1.29 & 1.17 & 6.63 \\
		$\delta_{\rm gal,obs}$ & 9.10 & 9.28 & 5.08 \\
		$C$ & 0.592 & 0.588 & 0.703 \\
		$V_{\rm true}$ [cMpc$^3$] & 6170 & 5630 & 16\,600 \\
		$\delta_{\rm m}$ & 1.84 & 1.86 & 1.21 \\
		$M_{z=0}~[10^{14}\,{\rm M}_\sun]$ & 7.14 & 6.57 & 14.9 \\
		$\delta_{\rm L}(z=0)$ & 2.56 & 2.55 & 2.08 \\
		$z_c$ & 0.80 & 0.79 & 0.38 \\
		\hline
		\multicolumn{4}{c}{Photometric Redshift Monte Carlo (Section \ref{sec:monte_carlo})} \\
		\hline
		$N_{\rm exp}$ & 					$1.87^{+0.94}_{-0.78}$ & 		$1.44^{+0.73}_{-0.68}$ & 		$5.59^{+1.56}_{-1.32}$ \\
		$\delta_{\rm gal,obs}$ & 				$6.30^{+5.91}_{-3.01}$ & 		$7.15^{+7.33}_{-3.34}$ & 		$3.44^{+1.62}_{-1.27}$ \\
		$C$ & 							$0.66^{+0.11}_{-0.13}$ & 		$0.64^{+0.11}_{-0.14}$ & 		$0.77^{+0.06}_{-0.06}$ \\
		$V_{\rm true}$ [cMpc$^3$] & 			$6630^{+1650}_{-950}$ & 	$5820^{+1630}_{-870}$ & 	$15\,600^{+1400}_{-1200}$ \\
		$\delta_{\rm m}$ & 					$1.42^{+0.80}_{-0.56}$ & 		$1.55^{+0.91}_{-0.59}$ & 		$0.89^{+0.32}_{-0.28}$ \\
		$M_{z=0}~[10^{14}\,{\rm M}_\sun]$ &	$6.54^{+4.34}_{-2.24}$ & 		$6.07^{+4.54}_{-2.10}$ & 		$12.1^{+3.3}_{-2.6}$ \\
		$\delta_{\rm L}(z=0)$ & 				$2.27^{+0.51}_{-0.56}$ & 		$2.35^{+0.52}_{-0.53}$ & 		$1.74^{+0.33}_{-0.38}$ \\
		$z_c$ & 							$0.55^{+0.43}_{-0.53}$ & 		$0.62^{+0.43}_{-0.48}$ & 		$0.06^{+0.32}_{-0.46}$ \\
		$P_{\rm collapse}$ & 				85.2\% & 					89.6\% & 					53.8\% \\
		\hline
	\end{tabularx}
	\end{center}
\end{table}

%%%%%%%%%%%%%%%%%
\subsection{Redshift of Collapse}
\label{sec:growth}
%%%%%%%%%%%%%%%%%

We showed in Section~\ref{sec:masses} that both overdensities have projected final masses $M_{z=0} \geq 10^{14}~{\rm M}_\sun$. To be protoclusters, each overdensity must also collapse and virialize by $z = 0$. We estimate the redshift of collapse analytically using the linear theory of density perturbation growth: we take the redshift of collapse, $z_c$, to be the redshift at which $\delta_{\rm L}$ exceeds the critical density, $\delta_c \simeq 1.686$, as has been demonstrated for several recently-detected $z \sim 2-3$ protoclusters \citep[e.g.,][]{Cucciati18, Polletta21}.

We calculate the linear overdensity $\delta_{\rm L}$ at the redshift of our overdensities as a function of $\delta_{\rm m}$ using the approximation in \citet{Mo96} Equation 18. We use Equation 7 from \citet{Cucciati18} to estimate $\delta_{\rm L}$ at lower redshifts, to find the redshift $z_c$ at which $\delta_{\rm L} = \delta_c$. For the photz-based overdensity calculations, we measure $z_c = 0.53$ in GNCL-z3.14 and $z_c = 0.61$ in GNCL-z3.10. This is consistent within our uncertainties (see Section \ref{sec:monte_carlo}) with the LF-based estimates of $z_c \approx 0.8$ for both overdensities. This indicates that they are indeed protoclusters and will collapse well before $z = 0$. These values, corresponding to lookback times of $5-7$ Gyr, are consistent with the detection of massive, virialized clusters in large Sunyaev-Zeldovich and optical-to-NIR surveys \citep[e.g.,][]{Bleem15, Balogh21}.

%%%%%%%%%%%%%%%%%%%%%%%%%%%
\subsection{Uncertainties and Statistical Significance}
\label{sec:monte_carlo}
%%%%%%%%%%%%%%%%%%%%%%%%%%%

\begin{figure*}[htb]
	\centering
	\includegraphics[width=0.9\linewidth]{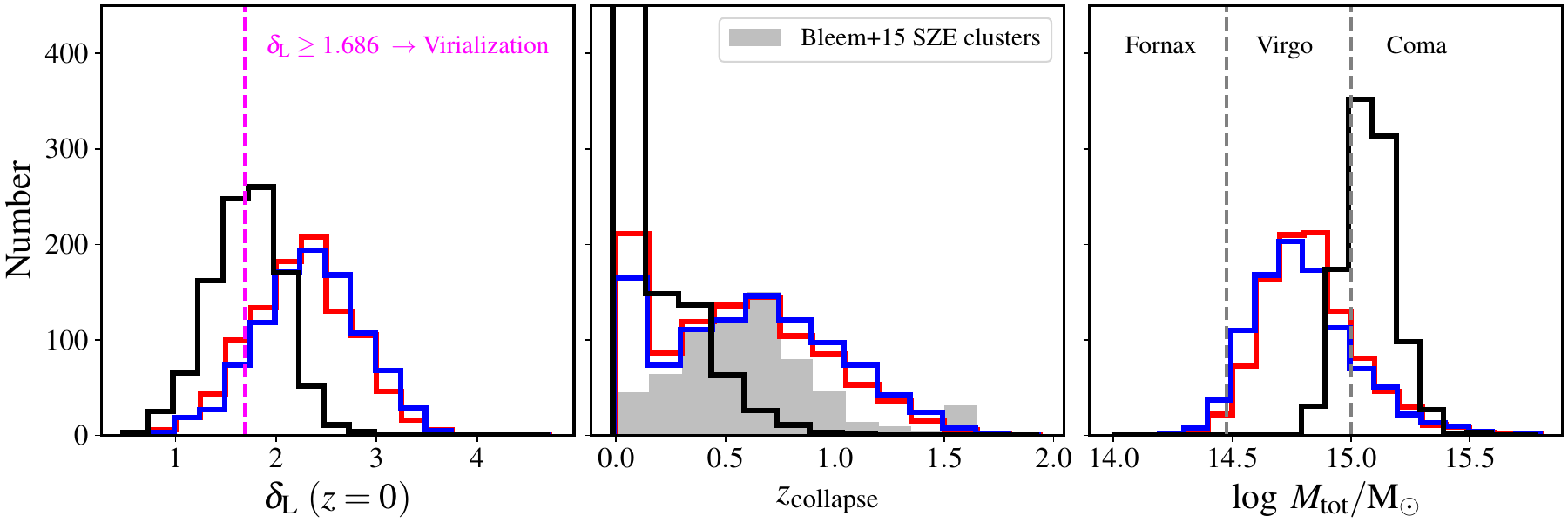}
	\caption{Histograms showing the results of our MC simulations of the protocluster overdensities, as described in Section \ref{sec:monte_carlo} and Appendix \ref{sec:mc_app}. The red, blue, and black histograms in each panel show the distribution of results for 1000 simulations of GNCL-z3.14, GNCL-z3.10, and the combined group, respectively. \textit{Left:} Linear overdensity at present day. The magenta dashed line denotes $\delta_{\rm L} = 1.686$, which is the threshold at which a spherical overdensity virializes. \textit{Center:} Redshift of collapse, i.e., the redshift at which the linear overdensity crosses the $\delta_{\rm L} = 0$ threshold. The first bin includes all simulations with $\delta_{\rm L}(z=0) < 1.686$, i.e., those that do not virialize by the present day. The grey histogram shows the redshift distribution of SPT Sunyaev-Zeldovich Effect-selected clusters from \citet{Bleem15}. \textit{Right:} Final ($z = 0$) masses of the cluster descendants. The grey lines at $M = 3 \times 10^{14}\,{\rm M_\sun}$ and $M = 10^{15}\,{\rm M_\sun}$ indicate the fiducial mass delineations between Fornax-, Virgo-, and Coma-mass clusters in the local universe \citep[e.g.,][]{Chiang13}.}
	\label{fig:outcomes}
\end{figure*}

To assess the statistical significance of these overdensities, we consider the likelihood that they would be observed and classified as protoclusters (i.e., have measured $M_{z=0} \geq 10^{14}~{\rm M}_\sun$), assuming the number of spectroscopic detections per bin follows Poisson statistics. To do so, we ran 1000 Monte Carlo (MC) simulations of the overdensity calculation described in Section~\ref{sec:photoz_overdensity} for each of the two overdensities. Instead of integrating $P(z_{\rm min} \leq z \leq z_{\rm max})$, for each iteration, we generated a random value of $z$ for every galaxy we targeted with Keck, and we summed these to measure a variable $N_{\rm exp}$. We describe this process in greater detail in Appendix~\ref{sec:photoz_pdf}. 

For each run, we calculated the associated galaxy and matter overdensities, redshift distortion, redshift of collapse, and $z = 0$ total mass in the same ways as in Sections \ref{sec:masses}--\ref{sec:growth}. In Table \ref{tab:protocluster_properties}, the quoted uncertainties are from the 16th to 84th percentile of these MC runs. Our MC simulations return 16th to 84th percentile final masses of $M_{z=0} = (4.3-10.9) \times 10^{14}\,{\rm M}_\sun$ and $M_{z=0} = (4.0-10.6) \times 10^{14}\,{\rm M}_\sun$ for GNCL-z3.10 and GNCL-z3.14, respectively. We show the histogram of mass outcomes for each of these in Figure \ref{fig:outcomes}. 

In all 1000 MC simulations for each overdensity, the projected $z = 0$ mass always exceeds $10^{14}~{\rm M}_\sun$. For some simulations with the lowest values of $\delta_{\rm gal,obs}$, we measure $z_c < 0$, indicating the overdense structure will not collapse by the present day and therefore cannot be considered bona fide protoclusters. We find that GNCL-z3.14 and GNCL-z3.10 do not collapse by $z = 0$ in 14.8\% and 10.4\% of our MC simulations, respectively. Once again, because these overdensities were performed using the photzs of our targets to calculate $N_{exp}$, these calculated likelihoods of collapse by the present day are underestimated. These MC simulations therefore suggest a $\gtrsim$85\% likelihood of collapse by $z = 0$.

We caution that this does not mean that GNCL-z3.14 and GNCL-z3.10 will remain separate until $z = 0$. Rather, the more likely scenario is that these two will collapse into substructures of a larger cluster with $M_{z=0} > 10^{15}\,{\rm M}_\sun$, which we address in Section \ref{sec:hyper}.

%%%%%%%%%%%%%%%%%%%%%%
\section{Substructures in a Hyper-Protocuster}
\label{sec:hyper}
%%%%%%%%%%%%%%%%%%%%%%
Recent deep spectroscopic surveys have detected so-called hyper-protoclusters: collections of galaxy overdensities and protoclusters in close proximity to each other \citep[e.g.,][]{Cucciati18}.
The two overdensities that we have identified in this work are separated by only 42 cMpc, assuming they have negligible peculiar velocities, and so may belong to such a structure.

Given their close proximity and high masses, GNCL-z3.14 and GNCL-z3.10 may be substructures in an extremely massive protocluster that will collapse into a Coma-like ($M > 10^{15}~{\rm M}_\sun$) cluster by $z = 0$. We can expect that these structures will collapse into a single halo at some point if the overall overdensity exceeds the critical density $\delta_{\rm L,crit} = 1.686$. To test whether this is possible, we repeat the overdensity and mass calculations described in Section \ref{sec:protoclusters} using a cylinder of radius 8~cMpc with faces at $z = 3.09$ and $z = 3.155$ and centered on the midpoint on the sky between the two protoclusters' coordinates. This new combined cylinder contains 25 spectroscopically detected galaxies.
In Table~\ref{tab:protocluster_properties}, we give the results. 

For both the LF- and the photz-based overdensity methods, we find that this structure will collapse before $z = 0$. We also repeat the MC procedure and measure a matter density of $\delta_{\rm L}(z=0) \geq 1.686$ in 53.8\% of runs, with very high projected $z = 0$ cluster masses of $M > 10^{15}~{\rm M}_\sun$. Furthermore, for the LF method, we calculate a matter overdensity of $\delta_{\rm m} > 1$ over co-moving volumes of $>16\,000$\,cMpc, which \citet{Chiang13} find to be consistent with a $>$80\% likelihood of collapse in simulated overdensities. 

To check this result, we calculate the likelihood that we have detected the progenitors of two separate clusters separated by 42\,cMpc.
Galaxy clusters and their progenitors are themselves subject to clustering over large scales \citep[e.g.,][]{Peebles80, Bahcall83}. This clustering is characterized by the correlation function, $\xi(r) = (r/r_0)^{-1.8}$ \citep{Totsuji69, Peebles80}, which is the excess probability of observing two objects (i.e., galaxies or clusters of galaxies) separated by a distance $r$ compared to an unclustered random Poisson distribution.

Given that GNCL-z3.14 and GNCL-z3.10 are the progenitors of either one (if they merge) or two (if they do not) clusters, Bayes' theorem gives the probability that the two overdensities will not merge by z = 0:

\begin{equation}
	P( 2 ~|~ 1~\mathrm{or}~2) = \frac{ P(2) }{ P(1~\mathrm{or}~2) } \, . 
\end{equation}
where N in each probability P(N) is the number of z = 0 clusters that are descendants of the overdensities we detect in our survey volume. We can rewrite this in terms of the Poisson probability and the correlation function:

\begin{equation}
	P(2 ~|~ 1 ~\mathrm{or}~ 2) = \frac{P'(2) \, [1 + \xi(r)]}{P'(1) + P'(2)\,[1 + \xi(r)]}  \label{eqn:prob_pair}
\end{equation}
where
\begin{equation}
	P'(N) = \frac{(nV)^N \, e^{-nV} }{N!} 
\end{equation}
is the Poisson probability of observing exactly $N$ objects within a volume $V$ given a number density of $n$.

The numerical value of the probability given in Equation \ref{eqn:prob_pair} depends on the number density $n$ and clustering length $r_0$, which differ greatly depending on the richness of the clusters considered. Since for either the single or dual descendant cases, the final masses of the protocluster(s) are very high ($> 5 \times 10^{14}\,{\rm M}_\sun$), we assume $r_0 = 25\,h^{-1}$\,cMpc and $n = 1.8 \times 10^{-6}\,{\rm cMpc}$, which were found for the richest and most massive clusters in deep SDSS cluster surveys \citep{Bahcall03}. Taking the volume to be our whole survey volume, roughly $20 \times 20 \times 870 \sim 3.5 \times 10^{5}\,{\rm cMpc}^3$ ($z = 2.8-3.8$), Equation \ref{eqn:prob_pair} gives a 34\% likelihood of GNCL-z3.10 and GNCL-z3.14 being progenitors of two distinct $z = 0$ clusters. This is likely an overestimation, as we prioritized objects in a narrower redshift range of $z = 3.0-3.3$ in our Keck survey, as noted in Section \ref{sec:obs}, which would yield a volume of $1.1\times10^5\,{\rm cMpc}^3$ and a two-descendant probability of only 14\%. We therefore claim an 86\% likelihood of the two structures merging.

Because the individual redshifts of collapse of GNCL-z3.10 and GNCL-z3.14 ($z_c^{\rm z3.10} \sim z_c^{\rm z3.14} \sim 0.5-0.8$) are higher than that of the combined Coma progenitor ($z_c^{\rm combined} \sim 0.1$ for the photz method and $z_c^{\rm combined} \sim 0.4$ for the LF method), the descendant of this system may be an interacting/merging cluster or cluster with multiple sub-clusters by $z \sim 0-0.4$. Several such clusters have been observed in this redshift range \citep[e.g.,][]{Clowe04, Jauzac16}. Indeed, a recent weak and strong gravitational lensing analysis shows that the rich $z = 0.308$ cluster Abell 2744 is composed of five mass peaks each with $M_{200} = (1-5) \times 10^{14}\,{\rm M}_\sun$ \citep{Cha24}, similar to the expected final masses of GNCL-z3.14 and GNCL-z3.10.

%%%%%%%%%%%%%%%%%%%%
\section{A Dusty Protocluster Core}
\label{sec:core}
%%%%%%%%%%%%%%%%%%%%
While both GNCL-z3.14 and GNCL-z3.10 have similar sizes, galaxy and matter overdensities, and expected total masses (see Section \ref{sec:protoclusters}), GNCL-z3.14 also has an excess of DSFGs, which may constitute a dusty protocluster ``core'' with an extremely high SFR density (SFRD), as seen in several other high-redshift protoclusters in the literature \citep[e.g.,][]{Miller18, Long20, Wang21}. While we leave a full analysis of the physical properties of the protocluster members to an upcoming companion paper (Nicandro Rosenthal et al., in prep.), we estimate the SFRs of the DSFGs belonging to GNCL-z3.14. To do so, we scale a \citet{Casey12} FIR spectral energy distribution (SED) with a fiducial dust temperature of $T_{\rm dust} = 32\,{\rm K}$ and emissivity $\beta = 1.8$ to the DSFGs' NOEMA 2\,mm continuum fluxes. We then obtain the integrated IR luminosities from these SEDs and convert these to SFRs following a \citet{Kennicutt12} law, corrected to a \citet{Chabrier03} initial mass function. The wavelengths at which we measured the continuum fluxes, integrated luminosities and SFRs inferred from this procedure are given in Table \ref{tab:sfr}. The DSFGs have individual SFRs of 300 to 1000 M$_\sun$\,yr$^{-1}$, and the total SFR of the four DSFGs is $\Sigma {\rm SFR} = 2730\pm670$.

\begin{table}
	\centering
	\caption{DSFG IR Luminosities and SFRs}
	\label{tab:sfr}
	\begin{tabularx}{\linewidth}{lcccc}
		\hline \hline
		Name & $\lambda_{\rm obs}$ & $S_{\rm 2mm}$ & $L_{\rm 3-1100\mu m}$ & SFR \\
		 & [mm] & [$\mu$Jy] & [$10^{12}\,{\rm L}_\sun$] & [M$_\sun$\,yr$^{-1}$] \\
		 \hline
		 GN-CL-2 & 1.895 & $870\pm30$ & $9.9\pm1.0$ & $990\pm100$ \\
		 GN-CL-3 & 1.895 & $900\pm30$ & $10.3\pm1.1$ & $1030\pm110$ \\
		 GN-CL-4N & 1.957 & $320\pm27$ & $4.0\pm0.5$ & $400\pm50$ \\
		 GN-CL-4S & 1.957 & $250\pm27$ & $3.1\pm0.5$ & $310\pm50$ \\
		 \hline
	\end{tabularx}
\end{table}

GNCL-z3.14 occupies a corrected co-moving volume of $6630^{+1650}_{-950}$\,cMpc$^{3}$. Accounting for only the DSFGs gives a lower limit of ${\rm SFRD} \geq 0.41^{+0.12}_{-0.14} ~{\rm M_\sun~yr^{-1}~cMpc^{-3}}$. If we instead use the volume of the combined massive protocluster of $15\,600^{+1400}_{-1200}\,{\rm cMpc^3}$, we find a lower limit of ${\rm SFRD} \geq 0.18 \pm 0.05 \,{\rm M_\sun\,yr^{-1}\,cMpc^{-3}}$. Both of these values are in line with known protoclusters in the literature \citep[see, e.g., Figure 11 of][and references therein]{Alberts22}.

Protocluster cores themselves typically have SFRs of several thousand ${\rm M_\sun\,yr^{-1}}$ and occupy only a few hundred cMpc$^3$ \citep[e.g.,][]{Miller18, Lacaille19, Long20, Wang21}. Following these studies, we can define a cylindrical volume around the four DSFGs with a radius of 2.6 cMpc and a depth of 13.3 cMpc, giving a total volume of 280 cMpc$^3$. In such a volume, the SFRD from the DSFGs alone is $9.7 \pm 2.4~{\rm M_\sun~yr^{-1}~cMpc^{-3}}$. This is also consistent with the densities of protocluster cores in the literature, which have been shown to have SFRDs of $\sim10-100\,{\rm M_\sun\,yr^{-1}\,cMpc^{-3}}$ \citep[e.g.,][]{Lacaille19, Wang21, Alberts22}.

Interestingly, apart from the presence of these DSFGs in GNCL-z3.14, both substructures identified in this work have very similar properties. They both extend over a similar sky area, displaying some apparent filamentary structure (as shown in Figure \ref{fig:3dmap}), both extend over a redshift range of $\Delta z \approx 0.02$, and both have very similar overdensities and therefore predicted final masses. The fact that no bright DSFGs are observed in GNCL-z3.10 could either be because of a true lack of DSFGs in this substructure, or because the DSFGs within it were too faint to cover with our survey. The former case is addressed by \citet{Miller15}, who find that many of the most dense volumes in the \textit{Bolshoi} cosmological simulations do not contain DSFGs with $S_{\rm 850} > 3\,{\rm mJy}$. While this trend is most pronounced at $z \lesssim 2.5$ where the most massive galaxies have stopped forming stars rapidly due to cosmic downsizing, \citet{Miller15} note that even at $z \gtrsim2.5$ the majority of the highest overdensities do not host a DSFG. 

Alternatively, there may be DSFGs in GNCL-z3.10 that are fainter than the SCUBA-2 selection for our NOEMA survey. As described in Table \ref{tab:dsfgs}, our NOEMA spectroscopic survey includes the six brightest galaxies in the Northwestern extension to the SCUBA-2 coverage of the GOODS-N (see also Figure \ref{fig:map}). This rules out the presence of the most highly star-forming Hyper-Luminous Infrared Galaxies (HyLIRGs; $L_{\rm IR} > 10^{13}\,{\rm L}_\sun$) like GN-CL-2 and GN-CL-3 in GNCL-z3.10, but not more moderately luminous DSFGs with $S_{850} = 4-6.5$\,mJy. \citet{Calvi23} leveraged deep photometric redshift data on the GOODS-N and found that DSFGs in this field do trace protoclusters, but the DSFGs occupying these protoclusters had a median $L_{\rm FIR} = 4 \times 10^{12}\,{\rm L}_\sun$, which would correspond to $S_{850} \approx 4\,{\rm mJy}$ at $z \sim 3$ for the dust SED described above. Additionally, deep interferometric observations have shown that some protoclusters have many fainter galaxies down to $<1$\,mJy that are still FIR-dominated and contribute significantly to the overall SFRD \citep[e.g.,][]{Hill20}.

Whether there are DSFGs in GNCL-z3.10 at a lower brightness or not, the contrast between the two structures warrants a further investigation of the properties of their member galaxies. While the absence of HyLIRGs in GNCL-z3.10 may not imply a difference in the mass distributions of these structures for the reasons stated above, it may indicate some difference in the evolutionary progression of the most massive galaxies in different components of the same larger protocluster, especially considering that there is evidence for correlated triggering of DSFGs in protoclusters \citep{Casey16}, and at least some DSFGs have very short lifetimes \citep[e.g.,][]{Carilli13}. 

Finally, because massive protoclusters such as the one presented in this work can span up to $30-60$ cMpc in the plane of the sky \citep[e.g.,][]{Tamura09, Cucciati18}, we inspect the speczs from \citet{Hsu19} of galaxies to the Southeast of our newly discovered protoclusters (i.e., the CANDELS field) to attempt to identify additional members, and especially additional DSFGs. We particularly aim to take advantage of the fact that \citet{Cowie17} have SMA positions for many galaxies in their SCUBA-2 sample, which allows us correctly identify NIR counterparts with existing speczs.

\begin{figure}[tb]
	\centering
	\includegraphics[width=\linewidth]{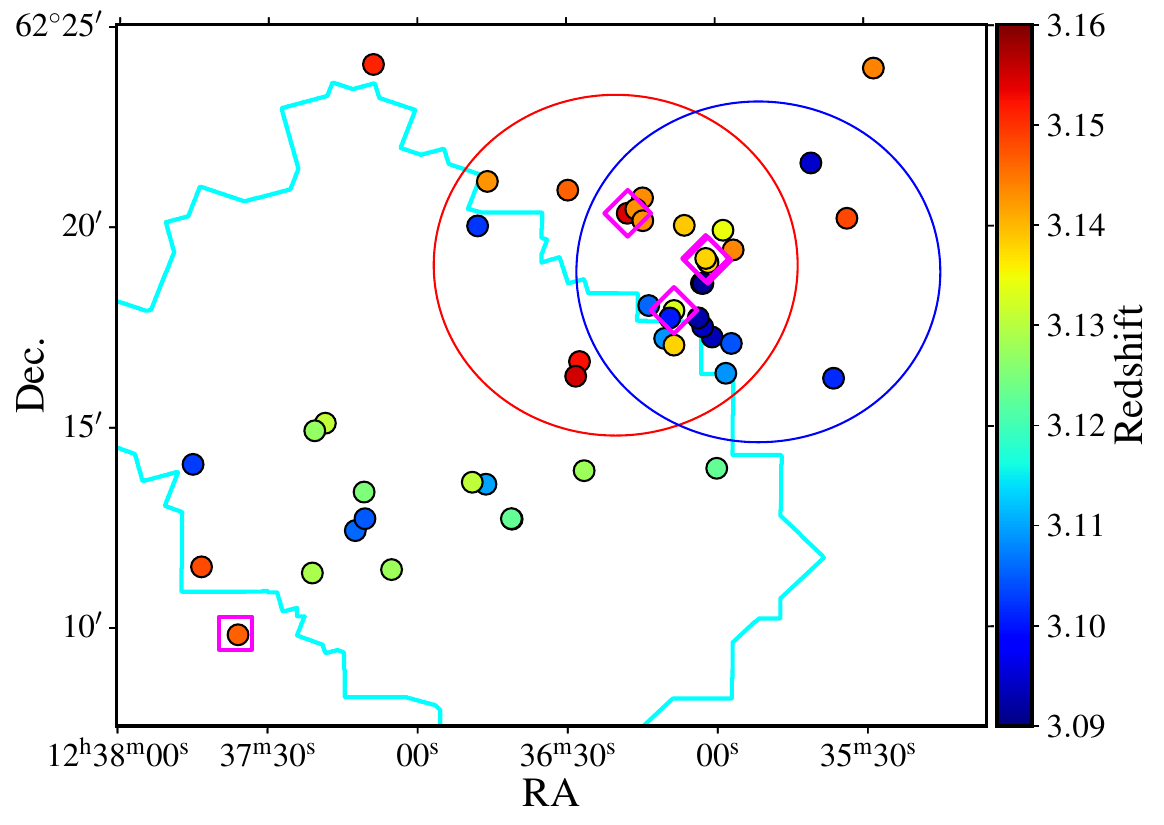}
	\caption{Positions of galaxies with $3.09 \leq z_{\rm spec} \leq 3.16$ in the GOODS-N, including both new speczs from this work and known speczs from the \citet{Hsu19} catalog. The cyan outline denotes the {\em HST}/WFC3 F160W coverage from CANDELS, and the large unfilled red and blue circles are the same as in the bottom-left panel of Figure \ref{fig:3dmap}. The inclusion of existing speczs from \citet{Hsu19} reveals a possible $z \approx 3.12$ substructure located to the Southeast of the two structures identified in Section \ref{sec:protoclusters}. The magenta diamonds indicate the $z_{\rm spec} = 3.133-3.155$ DSFGs GN-CL-2, 3, 4N, and 4S, and the magenta square shows the SMA position of an additional $z_{\rm spec} = 3.1465$ DSFG from \citet{Cowie17}.}
	\label{fig:fullmap}
\end{figure}

We identify 12 further objects with $3.09 \leq z_{\rm spec} \leq 3.16$ from the \citet{Hsu19}. One of these is the known NIR counterpart of a DSFG. This galaxy, shown by the magenta square in Figure \ref{fig:fullmap}, has a specz of $z_{\rm spec} = 3.1465$ and is a more moderately luminous galaxy with $S_{850} = 4.1$\,mJy \citep{Cowie17}. We show the positions of all galaxies in this redshift range in the GOODS-N, both with previously measured speczs and those with new speczs from this work, in Figure \ref{fig:fullmap}. The galaxies found to the Southeast are mostly at $z = 3.11 - 3.13$, lying between the redshifts of the two overdensities.

\section{Summary and Conclusions}
\label{sec:summary}

In this work, we presented spectroscopic observations of 507 galaxies just outside of the GOODS-N with the aim of definitively determining the presence or absence of a protocluster at $z \sim 3.14$ around two ultra-bright DSFGs. We measured 139 speczs, including NOEMA CO redshifts for four DSFGs, and confirmed the existence of two overdensities at $z = 3.090-3.110$ and $z=3.133-3.155$, which we refer to as GNCL-z3.10 and GNCL-z3.14, respectively.

We compared the number of spectroscopically confirmed members of each overdensity to the number of galaxies expected from their photzs and from the $K$-band LF of \citet{Mortlock17} at $2.75 \leq z < 3.75$. With both methods, we measured galaxy overdensities of $\delta_{\rm gal,obs} \sim 9$ over distortion-corrected volumes of $(5 - 7) \times 10^3~{\rm cMpc}^3$ for both overdensities. Using the analytical spherical collapse model of \citet{Mo96}, we calculated that both overdensities have matter overdensities of $\delta_{\rm m} = 1.6-1.9$ and expected $z = 0$ masses similar to that of the Virgo cluster ($M_{{\rm tot,}z=0} = (6-8) \times 10^{14}~{\rm M}_\sun$). We determined that both should collapse and virialize by $z_c \sim 0.5-0.8$.

We used the same overdensity calculation and spherical collapse analysis on a larger volume spanning $z = 3.09-3.16$ and found that the two overdensities combined have a galaxy overdensity of $\delta_{\rm gal,obs} \sim 3-5$ and matter overdensity of $\delta_m \sim 1$. We calculated a $>$80\% likelihood that GNCL-z3.14 and GNCL-z3.10 therefore are substructures in a single protocluster, rather than the progenitors of separate $z = 0$ clusters, and are expected to collapse into a single massive ($M_{tot,z=0} = (1.0-1.5) \times 10^{15}\,{\rm M}_\sun$) cluster, with a redshift of collapse of $z_c \sim 0.1-0.4$.

Lastly, while the two protoclusters are very similar in their estimated total masses and overdensities, GNCL-z3.14 contains four DSFGs with a cumulative SFR inferred from their submm emission of $\Sigma {\rm SFR} = 2730 \pm 670~{\rm M}_\sun~{\rm yr}^{-1}$ contained within a core volume of only 280~cMpc$^3$. These four DSFGs constitute a dusty protocluster core. The position of these two protoclusters just outside of the GOODS-N, in a region with extensive ground- and space-based photometric---and now spectroscopic---observations makes them prime candidates for follow-up investigations of their member galaxies, which may unveil whether there are significant differences in the two groups, and whether these are related to the lack of hyper-luminous DSFGs in GNCL-z3.10.

\begin{acknowledgements}

This work is based on observations carried out under project numbers S19CV, W19DG, and W20DH with the IRAM NOEMA Interferometer. IRAM is supported by INSU/CNRS (France), MPG (Germany) and IGN (Spain). M.J.N.R. and L.H.J. thank their IRAM contacts Vinod Arumugam, Isabella Cortez, and Melanie Krips for their valuable assistance in the reduction of these NOEMA data. We thank the anonymous referee for their helpful feedback.

We gratefully acknowledge support for this research from 
a Kellett Mid-Career Award and a WARF Named Professorship from the 
University of Wisconsin-Madison Office of the 
Vice Chancellor for Research and Graduate Education with funding from the 
Wisconsin Alumni Research Foundation (A.~J.~B.),
the William F. Vilas Estate (M.J.N.R., A.J.T.), and
NASA grant 80NSSC22K0483 (L.~L.~C.).

Some of the data presented herein were obtained at the W. M. Keck Observatory, which is operated as a scientific partnership among the California Institute of Technology, the University of California and the National Aeronautics and Space Administration. The Observatory was made possible by the generous financial support of the W. M. Keck Foundation.

The James Clerk Maxwell Telescope is operated by the East Asian Observatory on behalf of The National Astronomical Observatory of Japan; Academia Sinica Institute of Astronomy and Astrophysics; the Korea Astronomy and Space Science Institute; the National Astronomical Research Institute of Thailand; Center for Astronomical Mega-Science (as well as the National Key R\&D Program of China with No. 2017YFA0402700). Additional funding support is provided by the Science and Technology Facilities Council of the United Kingdom and participating universities and organizations in the United Kingdom and Canada.

The authors wish to recognize and acknowledge the very significant cultural role and reverence that the summit of Maunakea has always had within the indigenous Hawaiian community.  We are most fortunate to have the opportunity to conduct observations from this mountain. 

\end{acknowledgements}

\facilities{JCMT (SCUBA-2), NOEMA, Keck (LRIS, MOSFIRE)}

\software{\texttt{astropy} \citep{astropy13, astropy18, astropy22}, \texttt{eazy} \citep{Brammer08}, GILDAS (http://www.iram.fr/IRAMFR/GILDAS), \texttt{IDL} \citep{idl}, \texttt{matplotlib} \citep{Hunter07}, \texttt{numpy} \citep{Harris20}, \texttt{scipy} \citep{Virtanen20} }

\newpage

\appendix

\section{Calculating $N_{\rm exp}$ Using Our Photometric Selection}
\label{sec:photoz_pdf}

Calculating overdensities requires a calculation of the number of galaxies expected ($N_{\rm exp}$) within a given observed (i.e., not corrected for redshift distortion) volume. We selected our Keck targets based on their photzs from \texttt{eazy} \citep{Brammer08}, which were compiled by \citet{Hsu19} based on their UV-to-NIR photometry. \citet{Hsu19} also compiled speczs for a subset of these galaxies. To calculate $N_{\rm exp}$, we first define a parameter $q$ to be the difference between the specz and photz of a galaxy, normalized by one plus its photz:
\begin{equation}
	q \equiv \frac{z_{\rm spec} - z_{\rm phot}}{1 + z_{\rm phot}} \label{eqn:q} \,.
\end{equation}
Because every object in our survey has a fixed photz from \citet{Hsu19}, a PDF of $q$ can be used to draw random values of $z_{\rm spec}$ for each object, simply by inverting the equation above:
\begin{equation}
	z_{\rm spec} = z_{\rm phot} + q\,(1 + z_{\rm phot}) \,.
\end{equation}
While none of our targets had speczs prior to our survey, many objects from the \citet{Hsu19} sample have both photzs and speczs, and values of $q$ can be calculated for those objects. We use subsamples of these spectroscopically detected galaxies to generate the PDF of $q$ for each target.

For each individual object in our target sample, we select $N$ galaxies from \citet{Hsu19} that 1) have a specz, and 2) have a photz within 0.5 of the target object's photz. We then define the PDF $F(q)$ to be a sum of Gaussians, where each Gaussian is centered on $q_i$, which is the value of $q$ for the $i$th galaxy among the $N$ galaxy subsample of \citet{Hsu19}:
\begin{equation}
     F(q) = \frac{1}{\sqrt{2\pi \sigma^2}} \frac{1}{N} \sum_i^N \exp \left[ - \frac{(q - q_i)^2}{2\sigma^2} \right] \label{eqn:mixed_gauss_pdf} \,.
\end{equation}
In Equation \ref{eqn:mixed_gauss_pdf}, the Gaussian width $\sigma$ is the normalized median absolute deviation (NMAD) of the values of $q_i$ among the sample of $N$ galaxies, defined as
\begin{equation}
    \sigma \equiv 1.48 \times {\rm median} \left| q_i \right|_{i=1}^N \,.
\end{equation}

With our PDF defined, the probability of a given galaxy falling within $z_{\rm min} \leq z \leq z_{\rm max}$ is then 

\begin{equation}
    P(z_{\rm min} \leq z \leq z_{\rm max}) = \int_{q_{\rm min}}^{q_{\rm max}} F(q) \, T(q,z_{\rm phot}) \, dq \label{eqn:P(z)} \,,
\end{equation}
where the values of $q_{\rm min}$ and $q_{\rm max}$ are calculated from $z_{\rm min}$ and $z_{\rm max}$ from Equation \ref{eqn:q}. $T(q,z_{\rm phot})$ is a function of redshift that describes whether a galaxy would be detectable with MOSFIRE, based on the observability of strong lines (i.e., [O{\sc iii}]\,$\lambda$5008 and H$\alpha\,\lambda$6563) in the $K$-band. This factor is included so that $N_{\rm exp}$ is the number of galaxies that we expect to be \textit{detected} based on their photzs, such that we can accurately calculate $\delta_{\rm gal,obs}$ using our raw number counts of galaxies in each volume. For simplicity, we assume that $T(z)$ is proportional to the spectroscopic throughput of the telescope, instrument, and atmosphere at the observed wavelengths of these lines for any given redshift. From there, it is straightforward to convert $T(z)$ into $T(q,z_{\rm phot})$ using Equation \ref{eqn:q}.

We take $N_{\rm exp}$ to be the sum of these $P(z)$ for all galaxies we observed within a given sky area:
\begin{equation}
	N_{\rm exp} = \sum_j \, f_j\, P_j(z) \,,
\end{equation}
where $f_j$ is the MOSFIRE detection fraction for galaxies in our survey with $z_{{\rm phot},j} - 0.5 \leq z_{\rm phot} \leq z_{{\rm phot},j} + 0.5$.

\section{Monte Carlo Procedure}
\label{sec:mc_app}

To measure the uncertainties of the measurements produced by this procedure, we run MC simulations varying the values of $N_{\rm obs}$ and $N_{\rm exp}$ as follows:

We assume that $N_{\rm obs}$ follows Poisson statistics \citep[e.g.,][]{Gehrels86}. For each of the MC runs, we draw a value from the Poisson distribution assuming a mean of 13 for GNCL-z3.14, 11 for GNCL-3.10, and 25 for the combined group. These are the number of objects within the volumes outlined in Figure \ref{fig:3dmap} that also have photozs from \citet{Hsu19}. To also vary $N_{\rm exp}$, we draw a random redshift from the PDF given in Equation \ref{eqn:P(z)} for each observed galaxy within the $4.3'$-radius sky area. Instead of integrating the redshift PDFs for each galaxy, for the MC runs, we count each galaxy whose drawn redshift $z_j$ falls within the redshift range of the protocluster candidate, i.e.,
\begin{equation}
	N_{\rm exp} = \sum_{z_{\rm min} < z_j < z_{\rm max}}  f_j \,.
\end{equation}

We use this method rather than simply obtaining the posterior $P(z)$ distribution from \texttt{eazy}, as those distributions have a strong dependence on the choice of galaxy templates and the choice of priors used in their fits. Because we base our $P(z)$ function exclusively on the actual differences between $z_{\rm spec}$ and $z_{\rm phot}$ for galaxies fit identically to our targets, and $P(z)$ for an individual object is calculated based on the $\delta z$ of several hundred objects, this probability is agnostic to these biases.

{}

\end{document}